# Homogeneous Velocity-Distance Data
# for Peculiar Velocity Analysis.
# I. Calibration of Cluster Samples

Jeffrey A. Willick[1], Stéphane Courteau[2], Sandra M. Faber[3],
David Burstein[4], and Avishai Dekel[5]

[1] Carnegie Observatories, 813 Santa Barbara Street, Pasadena, CA 91101
[2] KPNO, P.O. Box 26732, Tucson, AZ 85726
[3] Lick Observatory, University of California, Santa Cruz, CA 95064
[4] Arizona State University, Dept. of Physics and Astronomy, Code 1504, Tempe, AZ 85287
[5] Racah Institute of Physics, The Hebrew University of Jerusalem, Jerusalem 91904, Israel

## ABSTRACT

We have combined five Tully-Fisher (TF) redshift-distance samples for peculiar velocity analysis: the cluster data of Han, Mould and coworkers (1991-93, HM) and Willick (1991, W91CL), and the field data of Aaronson *et al.* (1992), Willick (1991), Courteau & Faber (1992), and Mathewson *et al.* (1992), totaling over 3000 spiral galaxies. We treat the cluster data in this paper, which is the first of a series; in Paper II we treat the field TF samples. These data are to be combined with elliptical data (*e.g.*, Faber *et al.* 1989) to form the *Mark III Catalog of Galaxy Peculiar Velocities,* which we will present in Paper III. The catalog will be used as input for POTENT reconstruction of velocity and density fields, described in later papers, as well as for alternative velocity analyses.

Our main goal in Papers I & II is to place the TF data onto a self-consistent system by (i) applying a uniform set of corrections to the raw observables, (ii) determining the TF slopes and scatters separately for each sample, (iii) adjusting the TF zeropoints to ensure mutually consistent distances. The global zeropoint is set by the HM sample, chosen because of its depth and uniformity on the sky and its substantial overlap with each of the other samples.

In this paper, we calibrate the "forward" and "inverse" TF relations for HM and W91CL. We study the selection criteria for these samples and correct for the resultant statistical biases. The bias corrections are validated by comparing forward and inverse cluster distances. We find that many sample clusters are better modeled as "expanding" than relaxed, which significantly affects the TF calibrations. Proper corrections for internal extinction are derived self-consistently from the data.

*Subject headings:* Statistical Methods, Galaxy Distances, Cosmology

## 1. Introduction

An important approach to the problem of the origin and evolution of large-scale structure in the universe is the analysis of galaxy peculiar motions. Peculiar velocity studies are based on *redshift-distance samples,* which consist of radial velocity measurements ("redshifts") and redshift-independent distance estimates. While redshift measurements are straightforward and rarely suffer from systematic errors, redshift-independent distance determinations are complicated and are prone to such errors. Sytematic errors may occur *within* individual data sets, as a result of statistical biases which are not properly corrected for, or *between* data sets, as a result of zeropoint discrepancies among them. It is not possible at present to fully analyze the peculiar velocity field in the local universe ($cz \lesssim 6000$ km s$^{-1}$) using a single redshift-distance sample, as none adequately probes the entire volume. Peculiar velocity studies which merge redshift-distance samples must, however, take great care to ensure their uniformity.

In recent years large redshift-distance samples have been published by a number of independent groups, employing mainly the Tully-Fisher (TF) and $D_n$-$\sigma$ distance indicators for spiral and elliptical galaxies, respectively. Five major TF samples that have appeared over the last decade are of particular interest. The oldest of these data sets (Aaronson *et al.* 1982; hereafter A82) still provides the most complete information for over 350 nearby ($cz \lesssim 2000$ km s$^{-1}$) spirals. The A82 H band photometry has recently been reanalyzed by Tormen and Burstein (1995), eliminating small but significant systematic errors due to the use of a nonuniform set of diameters. Han, Mould and collaborators have published I band CCD TF data for 428 cluster spirals (Han 1991,1992; Han and Mould 1992; Mould *et al.* 1991,1993; hereafter, collectively, HM). The HM sample is unique among currently published TF samples in its depth and relatively uniform distribution on the sky. Willick (1991) has presented r band CCD TF data for 156 cluster galaxies distributed over the Northern sky, and for 385 field galaxies in the Perseus-Pisces region. Courteau and Faber (Courteau 1992; Courteau *et al.* 1993; hereafter CF) have presented an all-Northern sky TF sample of 325 galaxies, based on r band CCD photometry and optical velocity widths derived from H$\alpha$ rotation curves. The largest single contribution is that of Mathewson and collaborators (Mathewson *et al.* 1992; hereafter MAT), whose TF sample uses I band CCD magnitudes and a combination of H I and optical rotation widths for 1355 Southern Sky galaxies.

An unsatisfactory feature in the treatment of these data sets has been a lack of effort to ensure their homogeneity. The TF relation depends sensitively on the details of the measurements on which it is based. Photometric aperture and bandpass effects can affect the relation (Bothun and Mould 1987; Pierce and Tully 1988; Han 1991; Willick 1991), as can the choice of 21 cm vs. optical velocity width measurements (Courteau 1992). H I velocity widths themselves can differ systematically, depending on the algorithm used to determine profile width and on corrections for the effects of turbulence, resolution, and/or signal-to-noise ratio (*e.g.*, Bicay and Giovanelli 1987; Pierce and Tully 1988; Roth 1993). As a result of these different approaches, a multiplicity of TF relations actually exists; failure to match the TF relation to the data set can produce spurious results. Beyond the heterogeneity of the raw data, there has been no agreed upon method of *calibrating*—quantifying the relationship between luminosity and rotation velocity—the various TF relations. If sample selection effects could be neglected, the task would be a relatively simple one. But in the real world, selection effects make calibration of the TF relation (and its analogues such as $D_n$-$\sigma$) a difficult problem, as they are responsible for statistical biases that affect the calibration procedure (Willick 1994, hereafter W94).[1] It is largely in their treatment of selection bias that past approaches to distance indicator calibration have diverged.

A number of the present authors were involved in earlier attempts to merge and homogenize redshift-distance samples. The Mark I and Mark II Catalogs were privately distributed over email in 1987 and 1989 (respectively) by one of us (DB). These catalogs consisted primarily of the elliptical $D_n$-$\sigma$ data of Faber *et al.* (1989), Lucey and Carter (1988), and Dressler and Faber (1990), and spiral TF data published by the Aaronson group (A82; Bothun *et al.* 1985; Aaronson *et al.* 1989), supplemented by other data sets. The Mark II Catalog in particular has proven useful as a data base for peculiar velocity analyses, including early applications by some in our group of the POTENT technique for velocity and density field reconstruction (Dekel, Bertschinger, and Faber 1990; Bertschinger *et al.* 1990; Dekel *et al.* 1993). However, it is clear that with the advent of the large, new TF samples discussed above, the time is ripe for the construction of a more comprehensive catalog of homegeneous peculiar velocity data.

This paper is the first of a series in which our goal is to assemble and analyze such a catalog. In the present paper and the next (Willick *et al.* 1995b; Paper II), we present the basic methodology and results of the TF calibration for the five spiral samples. In the third paper of the series (Willick *et al.* 1995c; Paper III), we tabulate the basic TF data, the derived distances (including corrections for bias effects), and a large body of auxiliary information of potential interest; we also include the elliptical data from the Mark II compilation. The tabulated information in Paper III constitutes the *Mark III Catalog of Galaxy Peculiar Velocities*, which we will make available for electronic distribution as well. In the fourth (Faber *et al.* 1995) and fifth (Dekel *et al.* 1995) papers of this series, we will present a graphical description of the catalog and preliminary results of applying the POTENT algorithm to the Mark III data. We emphasize that although the initial application of the Mark III data will indeed be the POTENT analysis, the use of the Mark III Catalog is by no means restricted to POTENT. Rather, we hope that the Mark III will be considered a generally accessible data base useful for a variety of scientific purposes.

---

[1]W94 referred to the bias associated with sample selection effects as "calibration bias." Here we will use that term interchangeably with the intuitively simpler "selection bias."

In this paper, we treat the two Mark III TF samples that consist wholly of cluster galaxies: HM and W91CL. We pay special attention to deriving the TF zeropoint for the HM sample, as it will the determine the global zeropoint for the spiral samples (§ 2). In Paper II, we consider the Mark III spiral samples which do not resolve straighforwardly into clusters: the non-cluster portion of W91 (W91PP), and the CF, MAT, and A82 samples. The outline of this paper is as follows. In §2, we describe the general philosophy behind the TF calibration procedure, and the notation we will use in the TF and peculiar velocity analyses. In § 3, we describe the HM TF calibration; in § 4, the W91 TF calibration. In § 5, we present further tests and comparisons of these calibrations. In § 6 we summarize our main results.

## 2. Philosophy of Calibration and Notation

We emphasize at the outset a point previously made (*e.g.*, Lynden-Bell *et al.* 1988) but worth repeating. For peculiar velocity studies, it is not necessary to *absolutely* calibrate a distance indicator relation, *i.e.*, to obtain the estimated distances in physical units such as Mpc. It is sufficient that the relation yield distances in velocity units (specifically, $km\,s^{-1}$). Such "velocity distances" are defined as the product $r = HD$, where $H$ is the true Hubble constant and $D$ the true distance. One may thus meaningfully calibrate a distance indicator without determining either the Hubble constant or absolute distances separately, by working in a system of units in which only their product $r$ is relevant. It follows that the absolute magnitudes used in the distance indicator relation must be referenced not to a distance in Mpc, but to a distance in $km\,s^{-1}$. We define a galaxy's absolute magnitude as its apparent magnitude at a distance of $1\ km\,s^{-1}$. This unconventional definition greatly simplifies the transformation from observable quantities to distances in $km\,s^{-1}$.

To calibrate a distance indicator in such a system of units, we need models which yield distances to sample galaxies in $km\,s^{-1}$. These model distances (called "TF-orthogonal" by W94) are typically inferred from redshift-space information, *i.e.*, the measured radial velocity and position on the sky of a galaxy. Such information may be used in either of two distinct ways: in the exact sense of modelling distance as a mathematical function of redshift; or in the looser sense of identifying subsets of the overall sample whose individual members may be assumed to lie at the same distance. The latter type of model includes, but is not limited to, the well-known procedure of using cluster samples. The former approach generally requires a model of the peculiar velocity field—unless one makes the simplest assumption, namely, that peculiar velocities are negligible and objects obey pure Hubble flow.

Since our ultimate goal is to use calibrated distance indicator relations to reconstruct the peculiar velocity field (*e.g.*, using POTENT), independently of *a priori* assumptions, we will not base TF-orthogonal distances on peculiar velocity models. To do so would be to build into our TF calibrations prejudices regarding the specific nature of peculiar velocities—prejudices which might later manifest themselves in the velocity field we reconstruct. Throughout this paper and Paper II, TF-orthogonal distances will be based either on the group/cluster (common distance) assumption, on the assumption of pure Hubble flow (the "minimal" peculiar velocity model), or on some combination of the two. However, there is one sense in which we cannot remain entirely neutral with respect to the nature of peculiar velocities. Our definition of calibration requires us to "know"—in a suitable average sense—the distances in $km\,s^{-1}$ of at least a subset of sample objects. Otherwise, there is no way do derive a TF calibration in the specific sense discussed above.[2] Our only indicator of true distance in velocity units is, of course, the observed redshift. However, redshift is strictly equivalent to distance *only to the extent that the radial component of peculiar velocity vanishes*. This in turn implies that we can properly zeropoint the TF relation only for a sample characterized by a vanishing net radial peculiar motion.

To identify such a sample we need not assume that peculiar velocities are everywhere insignificant. But we must assume that small-scale peculiar velocities (flows into and out of highly over- and underdense regions) cancel when a large enough volume is considered, and that what remains on a large scale is at most a dipole pattern or "bulk flow." If a sample is located preferentially in one part of the sky, then a bulk flow will affect the mean radial peculiar velocities of sample objects. But if a sample is well distributed over the entire sky, the radial component of the bulk flow vector cancels in the sample as a whole. Of the five TF samples we treat, only one, the HM cluster sample, meets the criteria of large volume coverage and fairly uniform sky distribution. The TF relation for the HM sample is therefore the one which may be independently zeropointed. The TF zeropoints for the remaining samples will then be adjusted for statistical agreement with HM.

---

[2]More accurately, there is no way to derive a proper *zeropoint* for a TF relation without knowing distance in $km\,s^{-1}$. One can adequately determine the *slope* and *scatter* of the relation using arbitrary relative distances only.

Two other broad principles guide our construction of the Mark III Catalog. First, each TF sample is treated equivalently in terms of the corrections to the raw observables (mainly apparent magnitudes and velocity widths) which enter into the TF analysis. Thus, we do not adopt the corrected quantities published by the original authors, but instead take their raw quantities and submit them to our own corrections. We defer the details of these corrections to Paper III, but there is one issue we treat here and in Paper II: the value of the internal extinction coefficient. Following Burstein *et al.* (1995), we assume that the internal extinction correction is proportional to the logarithm of the major to minor axis ratio, and that the proper coefficient $C_{\rm int}$ may be determined by minimizing the TF scatter. Thus, determining $C_{\rm int}$ goes hand in hand with the TF calibration analysis. We assume that $C_{\rm int}$ depends only on photometric bandpass, and we estimate its value for the I, r, and H ($C_{\rm int}^{\rm I}$, $C_{\rm int}^{\rm r}$, and $C_{\rm int}^{\rm H}$) bands by minimizing the TF scatter of the HM+MAT, W91, and A82 samples respectively.

The second broad guideline is a uniform treatment of sample selection biases. Such effects arise because magnitude limits, diameter limits, or other observational restrictions distort the normal distributions needed for straightforward statistical analysis. It has been argued by some workers (*e.g.*, Federspiel, Sandage, and Tammann 1994) that these biases are so complex as to be amenable only to qualitative, graphical techniques. We agree that selection biases pose difficult problems, particularly because actual sample selection criteria are often somewhat murky. However, we have found that the selection criteria of the TF samples can be well enough approximated that the analytical bias formulae derived by W94 are applicable. In our treatment of each sample, we first characterize and quantify the selection criteria. Then, when we carry out the TF calibration analysis, we implement the iterative bias correction scheme developed by W94. (While it is not formally part of this series, the concepts and prescriptions of W94 will be invoked often enough that we recommend that the reader have some familiarity with that paper.) In order to test the validity of the bias corrections, we carry out calibrations not only of the "forward" but also of the "inverse" TF relations (these terms are defined in § 2.1). Selection biases are very significant for the forward relation, but are small and in some cases non-existent for the inverse. Through a suitable comparison of the two approaches, we can validate the bias-corrections, and hence the final forward calibrations. While we use the inverse TF calibration here and in Paper II mainly as a check on the forward, the inverse relation is valuable in its own right for peculiar velocity analyses, and we present inverse TF distances in Paper III.

## 2.1. Notation

In this series of papers we will adopt a coherent notation to describe the elements of a TF analysis; we summarize this notation in what follows. We parameterize the velocity width of a spiral galaxy by the symbol $\eta$, where by definition

$$\eta \equiv \log \Delta v - 2.5 \,; \tag{1}$$

here $\Delta v$ is the full velocity width (corrected for inclination) in $\mathrm{km\,s^{-1}}$. It is useful to think of $\Delta v$ as roughly twice the circular velocity of a spiral galaxy. However, $\Delta v$ may differ systematically from $2v_{\mathrm{rot}}$, depending on measurement technique. We at times refer to $\eta$ as the "velocity width parameter" or the "circular velocity parameter." We denote by $M$ the absolute magnitude of a galaxy. By "Tully-Fisher relation" we mean a mathematical function $M(\eta)$, whose precise meaning is

$$E(M|\eta) = M(\eta) \,, \tag{2}$$

where $E(x|y)$ signifies "expectation value of $x$, given $y$." The TF error is characterized by its rms dispersion (or "scatter") $\sigma$, defined by

$$E(M^2|\eta) - [E(M|\eta)]^2 = \sigma^2 \,. \tag{3}$$

We write the TF relation

$$M(\eta) = A - b\eta \,. \tag{4}$$

We refer to $A$ and $b$ as the TF "zeropoint" and "slope," respectively. The quantities $A$, $b$, and $\sigma$—the "TF parameters"—fully specify the TF relation. Each individual sample will have its own values of the TF parameters. (The validity of several assumptions implicit or explicit in Equations (2)–(4)—a gaussian distribution of $M$ about $M(\eta)$, $\sigma$ independent of $\eta$, and linearity of the TF relation—will be assessed in Paper III.)

We denote by $m$ the corrected apparent magnitude which enters into the TF relation. We distinguish the *raw* apparent magnitude with a subscript indicating photometric bandpass (*e.g.*, $m_I$ or $m_H$). Other galaxy data include the (blue bandpass) Galactic extinction $A_B$ and the logarithm of the (major to minor) axial ratio, which we denote $\mathcal{R}$. We refer to quantities $m$ and $\eta$ jointly as the "TF observables." We denote by $r$ a galaxy's true distance in $\mathrm{km\,s^{-1}}$, and define the corresponding distance modulus $\mu(r)$ without any normalizing constant, *i.e.*,

$$\mu(r) \equiv 5\log(r) \,. \tag{5}$$

*In the absence of all sample selection effects,* the operational definition of the TF relation is summarized by the equations

$$E(m|\eta, r) = M(\eta) + \mu(r) \,, \tag{6}$$

and

$$E(m^2|\eta, r) - [E(m|\eta, r)]^2 = \sigma^2. \qquad (7)$$

Equations (6) and (7) are the basis of the TF calibration.[3] In a real sample these equations are not satisfied, as a result of sample selection bias. W94 writes the actual expected apparent magnitude as

$$E(m|\eta, r) = M(\eta) + \mu(r) - \sigma\mathcal{B}(\eta, r). \qquad (8)$$

where the "relative bias" $\mathcal{B}$ is a function of sample selection criteria. An analogous expression describes the effect of bias on the observed scatter (see § 3.2).

The description above applies to the "forward" representation of the TF relation. It is in some cases preferable to view the relation in its "inverse" form, $\eta^0(M)$, with the specific meaning

$$E(\eta|M) = \eta^0(M), \qquad (9)$$

with scatter $\sigma_\eta$. The operational definition of the inverse TF relation consists of the equations

$$E(\eta|m, r) = \eta^0[m - \mu(r)] \qquad (10)$$

and

$$E(\eta^2|m, r) - [E(\eta|m, r)]^2 = \sigma_\eta^2, \qquad (11)$$

where again equations (10) and (11) are understood to hold in the absence of sample selection effects. We write the inverse relation

$$\eta^0(M) = -e(M - D); \qquad (12)$$

the quantities $e$ and $D$ are the "slope" and zeropoint of the inverse TF relation. It will not, in general, be the case that $D = A$ and $e = 1/b$; that is to say, the two representations of the TF relation are not mathematically inverse to one another (see § 3.3).

## 3.   The HM Cluster Sample

The HM sample forms the backbone of our calibration procedure. It consists of cluster spirals with I band CCD photometry and 21 cm spectroscopy published by Han, Mould, and collaborators: Mould *et al.* (1991; M91), Han (1992; HM92); Mould *et al.* (1993; M93). It is useful to divide the HM sample into North and South subsamples (according to the hemisphere from which the H I observations were done). We thus refer in what follows to "HM North," consisting of the ten Northern clusters in HM92 ("HM92 North") and all of M93, and "HM South," consisting of the six Southern clusters in HM92 ("HM92 South") and all of M91. Table 1 summarizes the overall characteristics of the HM sample, using our designations for the various subsamples. Column 1 gives the name for the cluster used by HM; Columns 2 and 3 give the mean galactic latitude ($l$) and latitude ($b$) of the cluster; Column 4 gives the CMB frame radial velocity in $\mathrm{km\,s^{-1}}$; Column 5 gives the number of galaxies used in the TF calibration analysis we present here. This number is not necessarily the same as the total number of galaxies with TF data, as we (like the original authors) have eliminated a number of galaxies from consideration on various grounds, such as dubious morphology, untrustworthy data, or redshifts inconsistent with cluster membership. In Paper III, we will provide data for all HM galaxies, and will indicate those not used in the TF calibration. The HM92 North subsample corresponds closely to W91CL (although W91CL contains an additional cluster, Ursa Major). In Column 6 of Table 1 we list, for future reference, the number of W91 galaxies in each of the HM92 North clusters used in the r band TF calibration analysis (§ 4.). We note, finally, that there exists still another HM subsample, a set of five clusters in the Perseus-Pisces region (Han and Mould 1992; HMPP). We choose not to include the HMPP subsample in our analysis here, since the Perseus-Pisces region of the sky is already probed by several of the HM92 North clusters (Pegasus, Pisces, A2634); inclusion of HMPP would overweight this region of the sky, possibly invalidating the utility of HM for setting the global TF zeropoint. However, we will present the HMPP data in Paper III, and will use this data when appropriate in the POTENT analysis.

---

[3]Equations (3) and (7) as written are not strictly consistent, as some additional variance is clearly introduced by photometric measurement errors. However, since Equation (3) is hypothetical only, as we never actually measure an absolute magnitude, we may take Equation (7) to be the operational definition of $\sigma$.

### 3.1. Selection Criteria Relations for HM

The statistical biases which affect distance indicator calibration may be corrected for, provided that sample selection criteria are suitably characterized (W94). Such criteria typically involve limits on the allowed magnitudes, diameters, or other apparent properties of sample objects. However, these limits in no case apply to the *CCD-measured* photometric properties of TF samples. Objects are never excluded based on their I band magnitudes (and of course they never fail to be detected in CCD photometry). Rather, the limits apply to "auxiliary" data, such as the diameters and magnitudes listed in photographic catalogs from which the candidates for TF observations were originally drawn. Further selection effects may then arise from the sensitivity limitations of 21 cm observations, which unlike the CCD photometry can result in nondetection. The selection criteria for the HM sample are unfortunately rather hybrid. In particular, they are quite different in the North as compared with the South. We focus first on the Northern Hemisphere clusters.

### 3.1.1. HM North

Only two galaxies in the HM North subsample are found neither the UGC (Nilson 1973) nor the Zwicky (Zwicky *et al.* 1961–68) catalogs. It is therefore reasonable to state that the HM North galaxies are limited in their optical properties by the inclusion criteria of the UGC and Zwicky catalogs. Two questions then arise: are the samples *complete* to the limits of these catalogs; and are there non-optical selection criteria which affect the makeup of these samples? While we cannot answer either of these questions with complete certainty, it is possible to characterize the sample selection procedure well enough for our purposes.

The UGC and Zwicky catalogs are limited by blue bandpass photographic apparent diameter ($D_B \geq 1'$) and magnitude ($m_B \leq 15.7$ mag), respectively. (It is useful to refer to the UGC diameter as $D_{\rm UGC}$, and to the Zwicky magnitude as $m_z$, leaving implicit that the measurement is made in a blue bandpass.) In Figure 1, we plot the UGC diameters and Zwicky magnitudes of sample galaxies as a function of log redshift. Several objects have diameters below the nominal UGC limit; there are a handful of such objects in the UGC catalog, but they do not affect the diameter limit substantially. Objects with $m_z > 15.7$ are UGC galaxies not in the Zwicky catalog. Galaxies with diameters less than 1' but otherwise unmeasured (not shown in the upper panel) are in Zwicky but not UGC. From Figure 1 we may draw two conclusions. First, many objects are at the catalog limits in diameter and magnitude. Second, at all redshifts sample objects tend to be concentrated near these limits. These are precisely the characteristics we would expect if indeed the samples are drawn to the limits of each catalog. By contrast, if the samples consisted typically of the brigther members of each cluster, we would expect that the lower envelopes of each figure would curve upward and to the left in each diagram. For HM92 North, there appears to be a perceptibly smaller degree of concentration near the catalog limits for redshifts $\lesssim 5000$ km s$^{-1}$. However, this latter effect is not strong, and does not preclude the existence of small or faint objects even at the lowest redshifts. We conclude that, insofar as the sample membership of HM92 North and M93 is constrained by the UGC and Zwicky inclusion criteria, these samples may be adequately described as being complete to the limiting diameter and magnitude of these catalogs.

We must also assess the possibility that H I nondetection might affect the makeup of HM

North. TF samples have often been described as "H I-selected"—indeed, this description has been used to argue against the need for consideration of sample selection bias (Aaronson *et al.* 1986). In Figure 2 the H I fluxes (upper panel) and fluxes per velocity channel (lower panel) are plotted as a function of log redshift. These two quantities are considered because it is possible to envision either one of them as limiting the sample. The distribution of these two quantities is quite different from that of the photographic magnitudes and diameters. At any given redshift, neither the fluxes nor the fluxes per channel are especially concentrated toward some limiting value. Moreover, the distributions are not constant with redshift; rather, the more nearby objects have, in the mean, larger H I intensities. This indicates that the samples do not probe to some limiting value; rather, they measure the typical H I properties of the sample as already defined by optical characteristics. This is not to say that H I detection is completely unimportant in the selection of these samples, only that that the samples are not strictly *limited* by their H I characteristics (this is not the case for the Southern HM samples, as we shall see). It is thus acceptable to neglect H I selection effects in our treatment of the HM northern Tully-Fisher samples. HM92 North and M93 will henceforth be described by what W94 refers to as "two-catalog selection": an object is in the sample if it satisfies *either* the UGC *or* the Zwicky catalog inclusion criterion.

The key to correcting the TF calibration procedure for selection bias is quantifying the relationships between the photographic quantities (in this case, $D_{\mathrm{UGC}}$ and $m_z$) on which selection is based, and the quantities which enter into the TF analysis. W94 treated the case that these relationships are adequately described by linear expressions of the form

$$\log D_{\mathrm{UGC}} \,=\, a_1 - b_1 m_I - c_1 \eta + d_1 \mathcal{R} - e_1 A_B \,, \tag{13}$$

and

$$m_z \,=\, a_2 + b_2 m_I + c_2 \eta + d_2 \mathcal{R} + e_2 A_B \,. \tag{14}$$

The negative signs in Equation (13) are adopted to ensure that all coefficients are positive. We denote the dispersions about these mean relations by $\sigma_\xi$ and $\sigma_z$, respectively.[4] The above relationships permit a derivation of sample selection probability *in terms of the TF observables*, the function W94 calls $S(m, \eta)$; see, for example, his Equations 32–36. This in turn leads to explicit expressions for the relative bias $\mathcal{B}$ for any given object, which is the essential ingredient in TF bias correction (which we describe in greater detail in § 3.2). In Equations (13) and (14) we have included terms involving axial ratio and Galactic extinction that do not appear in the simplified formulation of W94; they must be included in a treatment of real data. Moreover, we depart somewhat from the notation of W94 in that we use the *raw* rather than the corrected TF magnitude in the above equations. These modifications of the W94 formalism are easily incorporated into the TF bias correction procedure.

---

[4]Following W94, we often use the symbol $\xi$ to represent $\log D$, where $D$ is expressed in units of $0.1'$; in this convention $1'$ corresponds to $\xi = 1$.

The coefficients in Equations (13) and (14) may be solved for by means of a multiparameter least-squares fit. However, such a fit, when carried out in a single step, will yield biased results because the dependent variables $\log D_{UGC}$ and $m_z$ are each subject to strict limits ($\xi \geq \xi_\ell = 1$ and $m_z \leq m_\ell = 15.7$). Consequently, the iterative correction procedure developed by W94 must be applied to the determination of the coefficients in Equations (13) and (14), just as we will later apply it to the TF calibration. In the present case, the applicable formulae are those describing a *strict* magnitude or diameter limit (W94 § 3). Specifically, an initial fit yielded estimates of the coefficients and scatter; this in turn permitted an estimate of the biases in $\xi$ and $m_z$; the observed values of $\xi$ and $m_z$ were then "corrected" by these bias values, and the fit redone. The least-squares scatter estimates (which are biased low) were also corrected. The fit was iterated until convergence, as measured by the fitted coefficient values, was achieved.

This iterative fit procedure was applied to the 145 objects in HM North with UGC diameters greater than $1'$. The resulting coefficients and scatter are given in the first line of Table 2. The coefficient of Galactic extinction $A_B$ is identically zero, because the value of this coefficient was found to be statistically consistent with zero, and the fit was redone with $A_B$ not considered. In the top panel of Figure 3, we plot the UGC diameters, fully corrected for bias at the end of the iteration procedure, as a function of the I band magnitudes. To the extent that the bias-correction procedure has been successful, the plot depicts the "true" relationship between $D_{UGC}$ and the TF observables, seen as it would be in the absence of a diameter limit. As expected, many of the corrected diameters are now below the catalog limit of $1'$. An analogous fit using the Zwicky magnitudes was applied to the 187 objects in HM North with $m_z \leq 15.7$ mag, with results given in the second line of Table 2. It is perhaps surprising to note that the coefficient of the Galactic extinction was again found to be negligible. Although it is likely that a dependence of $m_z$ on $A_B$ exists in reality, it could not be detected with the limited data available here. In the bottom panel of Figure 3 we plot the bias-corrected Zwicky magnitudes vs. the raw CCD I band magnitudes. As in the case of the UGC diameters, the corrected photographic magnitudes can be consderably fainter than the Zwicky catalog limit (shown as a dotted line). The fact that the statistically expected UGC diameters and/or Zwicky magnitudes fall below the catalog limits for many sample objects is the source of the TF calibration bias we discuss in detail in § 3.2. We note, finally, that in the fits for $\log D_{UGC}$ and $m_z$ we searched for, but did not detect, a redshift-dependence of the fit residuals. We also tested whether including a term linear in $\log(cz)$ improved the fit, and it did not.

### 3.1.2. HM South

The selection of TF candidate galaxies for HM South was made from the ESO (Lauberts 1982) catalog. ESO is diameter limited by the criterion $D_B \geq 1'$ (the ESO and UGC visual diameters are not necessarily on the same system, so these catalogs are not necessarily complete to the same depth). As shown in Figure 4, where ESO diameters are plotted against log redshift, HM South galaxies do not satisfy a clear, uniform diameter limit. At the lowest redshifts ($cz \lesssim 2500\,\mathrm{km\,s^{-1}}$), the limiting diameter, to the extent one exists, is in the range 2–2.5'. At the highest redshifts, $cz \gtrsim 4000\,\mathrm{km\,s^{-1}}$, the M91 sample approaches, but does not quite reach, the ESO catalog limit; HM92 South galaxies typically have diameters $\gtrsim 1.5'$ at all redshifts.

These statements are unfortunately not very precise, but they reflect the nonuniform selection criteria of the HM South samples. Neither M91 nor HM92 South show the strong clustering toward a single diameter limit exhibited by their northern counterparts. That this is the case is due in part to the importance of neutral hydrogen-related selection effects in HM South, depicted in Figures 5 and 6. In Figure 5 we plot the HM South H I fluxes, and fluxes per velocity channel, as a function of log redshift. The latter quantity is an approximate measure of signal-to-noise, since the baseline noise in the H I spectra do not vary very much. Although the plots are similar, it seems more likely that the flux-per-channel, rather than the flux itself, would limit the sample, and we restrict our attention to it in the following. We see that, at least for redshifts $\gtrsim 2500\,\mathrm{km\,s^{-1}}$, the fluxes per channel do tend to have a limiting value of $\sim$ -2 at all redshifts. This contrasts with the situation in the North in two ways: first, the fluxes per channel do not tend to diminish with higher redshift. Second, the typical fluxes per channel are higher by a factor of $\sim 3$ than the corresponding values in the North. The probable reason for this is the great sensitivity of the Arecibo radio telescope in the North, as compared with Parkes in the South. Moreover, HM92 south and M91 are quite similar in their H I properties (whereas their ESO diameter distribution was markedly different), suggesting that the 21 cm data played an important role in sample selection. In Figure 6, we plot the H I fluxes per channel versus log $D_{\mathrm{ESO}}$. The distribution in this plane has the character of being limited in *both* coordinates. This further suggests that the sample is limited by H I flux per channel as well as by the ESO optical diameter.

While these considerations are not entirely conclusive, we adopt the following characterization of HM South: We suppose that this subsample is limited both by ESO diameter and by 21 cm "signal-to-noise," as embodied in the flux per channel parameter $\log(F_{\mathrm{H\,I}}/\Delta v)$ plotted in Figures 5 and 6. These limits are not in the sense "either-or," as was the case for HM North. Rather, the dual selection criteria in the South are in the "both-and" sense: objects must be both large enough (as measured by their ESO diameter) *and* sufficiently intense in 21 cm to be included (W94 has analyzed the bias effects arising for both types of selection). It is evident from the above that the limiting values of both $D_{\mathrm{ESO}}$ and $\log(F_{\mathrm{H\,I}}/\Delta v)$ are not well defined. When we carry out the HM TF calibration we will adopt limiting values of $D_{\mathrm{ESO}} = 1.8'$ and $1.6'$ for HM92 South and M91, respectively, and $\log(F_{\mathrm{H\,I}}/\Delta v) = -1.9$ for both subsamples. These choices represent compromises between the *absolute* limits on the diameters and fluxes-per-channel, which are clearly fainter, and the fact that there is insufficient clustering near the absolute limits to treat them as true limits. The necessity of such a compromise is unsatisfying, but we will show in § 5.1. that the calibration bias corrections which depend on the chosen limits are sufficiently accurate for our purposes. An alternative to adopting compromise values of the diameter limits would be to model the limits as "fuzzy." Such an approach is mathematically feasible, but would introduce addtional, and unwarranted, complexity into the calibration analysis.

We characterize the $\log D_{\mathrm{ESO}}$–$m_I$ relationship in analogy with the method used for the UGC diameter and HM North. We use the 103 HM South galaxies with $D_{\mathrm{ESO}} \geq 1.5'$ in the fit.[5] As before, we correct for the bias due to this diameter limit. When an initial fit of $\log D_{\mathrm{ESO}}$ to $m_I$ and $\mathcal{R}$ only was carried out, the results were found to differ qualitatively from those of the corresponding fit of the UGC diameters, as is shown in Figure 7. The left hand panel shows the $\log D_{\mathrm{ESO}}$–$m_I$ relation, while the two right hand panels show the residuals from this fit with respect to $\eta$ and $\log$ redshift. As can be seen, the residuals do not correlate with $\eta$, but do correlate with $\log(cz)$. To account for this observation we now fit a relation of the form

$$\log D_{\mathrm{ESO}} \; = \; a_1 - b_1 m_I - c_1[\log(cz) - 3.5] + d_1 \mathcal{R} \, . \tag{15}$$

The coefficients resulting from this fit are given in Table 2. With the addition of the redshift term the fit improved significantly, the scatter falling from 0.124 to 0.112 dex. Along with the redshift dependence goes a markedly smaller coefficient of $m_I$ than was seen in the UGC diameter fit to the HM92 North subsample. It is not clear why the ESO diameters should differ in these ways from the UGC diameters in their relation to the CCD magnitudes; as we shall see in Paper II, the $\log D_{\mathrm{ESO}}$–$m_I$ relation derived from the Mathewson data is very similar. A possible explanation of the redshift-dependence of the diameter-magnitude relation and its consequences for *Malmquist* bias correction are discussed by Willick (1995). For our current purpose of correcting the TF calibration for *selection* bias, this redshift-dependent $\log D_{\mathrm{ESO}}$–$m_I$ relation poses no special problems.

---

[5]This may be an underestimate of the actual limit—compare with the larger limiting values of $\log D_{\mathrm{ESO}}$ mentioned in the previous paragraph—but a much larger limiting value significantly reduces the number of objects participating in the fit.

To characterize H I-related selection effects we fit the H I fluxes to the I band apparent magnitudes. However, in a break with our usual convention, we have used not the raw, but rather the fully corrected I band magnitudes. This is because there is little reason to think the H I fluxes should be in any way affected by internal or Galactic extinction. We found when we carried out a two-parameter linear fit of $F_{H I}$ to $m$ that there was again a correlation of the fit residual with redshift. This correlation is exhibited in the upper panel of Figure 8. The physical origin of this correlation remains unclear. A second fit was carried out with an additional term in log redshift included; as in Equation (15), this term was normalized at $\log(cz) = 3.5$, and carried a minus sign. The resultant parameters from this fit, including its scatter, are shown in Table 2. The lower panel of Figure 8 shows that the inclusion of the redshift-dependent term has eliminated any trend of the residuals with redshift. Further possible dependencies of the H I flux—on $\eta$, axial ratio, and Galactic extinction—were tested for and found to be negligible.

### 3.2. Calibration of the HM Tully-Fisher Relation

Our calibrating sample consists of the 346 galaxies in 31 clusters listed in Table 1. We assume at first that within any single cluster, the individual galaxies lie at a common distance from us (the "cluster paradigm"). We refer to the distance modulus of the $i^{th}$ cluster as $\mu_i$, and to the TF observables of the $j^{th}$ galaxy in the $i^{th}$ cluster as $(m_{ij}, \eta_{ij})$. Neglecting for the moment sample selection effects, we then have (Equation 6):

$$E(m_{ij}|\eta_{ij}) = M(\eta_{ij}) + \mu_i. \qquad (16)$$

Note that the $\mu_i$ enter into the analysis at this point as free parameters. Their best-fitting values are to be determined through a statistical analysis which uses the TF information, but only in conjunction with the strong constraint imposed by our adoption of the cluster paradigm. Thus, although unknown at the outset, the $\mu_i$' constitute TF-orthogonal distance estimates in the sense of § 2; our analysis is therefore subject to selection bias (W94).

We could in principle replace $M(\eta_{ij})$ in the above equation with $A - b\eta_{ij}$, according to Equation (4). However, doing so would assume that our cluster distances are expressed in km s$^{-1}$, whereas at present our distance scale is entirely arbitrary. We thus combine the TF zeropoint $A$ and the cluster distance moduli $\mu_i$ into a single constant $a_i$, giving

$$E(m_{ij}|\eta_{ij}) = a_i - b\eta_{ij}. \qquad (17)$$

The $a_i$'s which appear in Equation (17) are temporary devices, to be discarded later; however, the quantity $b$ is the true slope of the HM TF.

To the extent that each galaxy apparent magnitude is identically and independently distributed about its expectation value, Equation (17) suggests that we determine the unknown free parameters $b$ and the $a_i$ by minimizing a sum-of-squares of the form

$$\mathcal{S} = \sum_{i=1}^{N} \sum_{j=1}^{n_i} w_i \left[ m_{ij} - (a_i - b\eta_{ij}) \right]^2 , \tag{18}$$

where $N$ is the number of clusters and $n_i$ the number of galaxies in the $i^{th}$ cluster. The $w_i$ are weights which can accentuate our intuitive sense that the more objects there are in a cluster, the more significant a role that cluster should play in determining the TF slope. We have chosen a weighting function of the form

$$w_i = 1 - e^{-\frac{n_i}{n_0}} \tag{19}$$

and have adopted $n_0 = 4$ (experimenting with different values shows the slope in fact to be relatively insensitive to the precise value of $n_0$). The very small cluster samples contribute in smaller proportion than their numbers; cluster samples with $n_i \gtrsim 10$ are weighted essentially in proportion to their size. Minimizing $\mathcal{S}$ in Equation (18) results in a system of $N + 1$ coupled linear equations involving $b$ and the $a_i$, which may be solved using standard techniques.

However, a single-step solution of this system of equations yields biased results. The ultimate source of this bias is that only a fraction $\leq 1$ of all galaxies in a cluster meet the sample selection criteria and thus participate in the calibration analysis. Because of the correlations summarized in Table 2, the "missing" galaxies, at a given $\eta$, are likely to be fainter (in TF magnitude) than those found in the sample; the latter, as a result, are brighter in the mean than the TF prediction $a - b\eta$. At a given distance, this effect increases with decreasing $\eta$; at a given $\eta$, this effect increases with distance. Two immediate consequences are that the apparent TF relation is too flat and the relative distance moduli are too low, with the latter effect greater for the distant clusters. In addition to these obvious consequences, the apparent TF scatter is too small, because objects tend to "pile up" near the sample limits and thus not exhibit their true dispersion. In order to correct for these effects, we follow the prescription of W94:

1. Carry out the fit once, neglecting all bias effects. Obtain preliminary estimates of $b$, $\sigma$, and the $a_i$.

2. Use this preliminary fit, along with the sample selection relations summarized in Table 2, to estimate, the various bias parameters defined by W94. These parameters are of two sorts: the first are "coupling parameters" (called $\alpha$ and $\beta$) which measure the relative scatter of the sample selection relations as compared with the TF relation and are constants for any subsample; the second are the magnitude and diameter limit "closeness parameters" (called $\mathcal{A}_\xi$ and $\mathcal{A}_z$), functions of $\eta$ and thus different for each object, which measure the effective closeness to the sample limits. (See Equations 40, 41, 65, and 66 of W94.)

3. For each galaxy, compute the relative bias $\mathcal{B}$, which is a function of the bias parameters $\alpha$, $\beta$, $\mathcal{A}_\xi$, and $\mathcal{A}_z$. The specific form of the function depends on whether the subsample in question is described by two catalog (either-or) selection (for HM North; see W94 Equation 67) or two criteria "both-and" selection (for HM South; W94 Equation 77).

4. Correct the observed TF magnitudes according to

$$m^{(c)} = m + \sigma \mathcal{B} . \tag{20}$$

5. Re-solve the system of equations, now using the corrected apparent magnitudes $m^{(c)}$ rather than the observed apparent magnitudes $m$, for $b$ and the $a_i$.

6. Redetermine the scatter $\sigma$, now correcting for the *dispersion bias* described by W94. This bias is quantified in terms of a "relative variance" $\mathcal{C}$, which is also a function of the parameters $\alpha$, $\beta$, $\mathcal{A}_\xi$, and $\mathcal{A}_z$. The corrected scatter estimate is given by

$$\sigma^2 = \frac{1}{N_{\rm tot} - p} \sum_{i=1}^{N} \sum_{j=1}^{n_i} \frac{[m_{ij}^{(c)} - (a_i - b\eta_{ij})]^2}{\mathcal{C}} , \tag{21}$$

where $N_{\rm tot}$ is the total number of galaxies participating in the fit, and $p = N + 1$ is the number of free parameters in the model.

7. Use the new values of $b$, the $a_i$, and $\sigma$ to again calculate the bias parameters as in step 2. Proceed as before through steps 3–6.

8. Iterate this procedure until convergence, as indicated by constancy of the TF parameters, is achieved.

The HM TF parameters resulting from carrying out this procedure are given in the first line of Table 3. We discuss below the determination of the TF zeropoint $A$, which we have not touched upon thus far. However, first we turn to a significant departure from the basic model, which in fact renders these preliminary results inadequate.

### 3.2.1. Departure from the Cluster Paradigm

Up to this point, we have invoked the cluster paradigm based on nothing more than the fact that the sample objects were identified as "cluster galaxies" according to certain redshift-space criteria adopted by the original authors. But the common-distance assumption is at best an approximation, since even a dense, virialized cluster has some physical depth. Moreover, it is far from clear that the spiral galaxies which make up our sample typically occupy the virialized cores of cluster. Fortunately, the results of the TF calibration enable us to test the validity of the cluster paradigm for each cluster in the sample. To the extent that it is an adequate approximation, the TF residuals should exhibit no detectable correlation with redshift. Departures from the cluster paradigm should, by contrast, manifest themselves as TF residual–radial velocity correlations.

In Figure 9, we plot TF residuals (*i.e.*, observed minus predicted apparent magnitude, following bias correction) vs. log redshift, for nine clusters in the HM sample. We have chosen these nine clusters because of the (often very pronounced) correlations of the TF residuals with redshift in these clusters. The remaining twenty-two clusters in the sample, when plotted in a similar fashion, do not manifest a similar trend. In each panel in the figure, we show the straight line, of slope 5, which passes through the median redshift of the cluster and residual zero. We would expect that the residuals would fall approximately along such a line if, rather than lying at a common distance, the individual cluster galaxies roughly followed the Hubble expansion. The approximate adherence of the residuals to a line of slope 5 in these clusters leads us to consider a model which accounts for the apparent redshift-distance trend, which we construct as follows.

We suppose that in each of the nine apparently "expanding" clusters, the distance of a given galaxy from the cluster center is proportional to its radial velocity difference from the cluster mean. Thus, if cluster $i$ has mean distance $d_i$ and redshift $v_i$, the $j^{th}$ galaxy in that cluster lies at a distance

$$d_{ij} = d_i + H_i^{-1}(v_{ij} - v_i),\qquad(22)$$

where $H_i$ is the local expansion rate around the cluster. To fit the model indicated by Equation (22) would involve solution for an additional nine free parameters, *i.e.*, the values of $H_i$ for each of the expanding clusters. The data do not warrant this, however; Figure 9 suggests that the slopes of the residual-log(redshift) correlations are not well constrained in general. To simplify we assume that the local expansion rate for each cluster is equal to the global expansion rate, which is unity in our system; thus

$$\frac{d_{ij}}{d_i} = 1 + \frac{v_{ij} - v_i}{d_i}.\qquad(23)$$

If we now write $d_i = v_i - u_i$, where $u_i$ is the radial component of the cluster peculiar velocity, and make the reasonable assumption $|u_i/v_i| \ll 1$, Equation (23) may be approximated by

$$\frac{d_{ij}}{d_i} = 1 + \frac{v_{ij} - v_i}{v_i} + \frac{(v_{ij} - v_i)u_i}{v_i^2}.\tag{24}$$

where we have expanded to second order in the small quantity $u_i/v_i$. The last term on the right hand side of Equation (24) will typically be of order $10^{-2}$, since both velocity differences from the cluster mean and peculiar velocities are measured in the hundreds of $\mathrm{km\,s^{-1}}$, whereas cluster radial velocities are measured in the thousands of $\mathrm{km\,s^{-1}}$. In addition to being small, the last term will be as often positive as it is negative in any given cluster. Therefore, it acts only as a small additional variance in the model but has no systematic effect, and we may neglect it. Doing so and taking logarithms we obtain

$$\mu_{ij} = \mu_i + 5\log\left(\frac{v_{ij}}{v_i}\right).\tag{25}$$

Note that while Equation (25) describes an "expanding" cluster, it does *not* require that $\mu_i = \log v_i$; it does not, in other words, describe "pure" Hubble flow.

Using Equation (25), we rewrite Equation (17), the expected apparent magnitude of the $j^{th}$ galaxy in the $i^{th}$ cluster, as

$$E(m_{ij}|\eta_{ij}) = s_i \times 5\log\left(\frac{v_{ij}}{v_i}\right) + a_i - b\eta_{ij},\tag{26}$$

where the $s_i = 1$ or $s_i = 0$, depending upon whether cluster $i$ is or is not treated as "expanding." The $a_i$ retain their earlier meaning as arbitrary relative distance moduli. The TF calibration is now based on minimization of the following suitably modified sum-of-squares term:

$$\mathcal{S} = \sum_{i=1}^{N}\sum_{j=1}^{n_i} w_i\left[\left(m_{ij} - s_i \times 5\log\left(\frac{v_{ij}}{v_i}\right)\right) - (a_i - b\eta_{ij})\right]^2.\tag{27}$$

Using this model we obtain the TF parameters given in the second line of Table 3. The expanding cluster model has reduced the overall TF scatter from $\sim 0.46$ to $\sim 0.39$ mag. The formal statistical uncertainty in the scatter estimate is given by $\sigma/\sqrt{2(N-p)}$, where $p$ is the number of free parameters. Taking $\sigma = 0.4$ gives a formal uncertainty $\delta\sigma = 0.016$, which indicates that the reduction in the scatter associated with the adoption of the expansion model is highly significant. We also show in parentheses the 1-$\sigma$ uncertainties in the TF zeropoint and slope. These uncertainties include only random statistical effects; they do not include the additional uncertainty associated with application of the bias-correction algorithm (see W94 for a further discussion of this issue). We note also that the expansion model has had a substantial effect on the zeropoint (whose calculation is discussed in § 3.2.2) as well as the scatter. This arises because of the effect of the reduced TF scatter on the bias corrections. In Figure 10, we plot new TF residuals vs. log redshift for the same nine clusters shown in Figure 9. As can be seen, the trends with redshift evident in the earlier plots have been essentially eliminated. Finally, in the third line of Table 3 we show for reference the parameters resulting from the initial minimization of the sum-of-squares in Equation (27), prior to application of the bias correction procedure. Note that all of the TF parameters from this uncorrected fit differ markedly from their corrected values. The reasons for these differences are discussed in detail by W94; we further discuss the signficance of the corrections in § 5.

There remains a subtle point of statistical analysis, associated with the adoption of the cluster expansion model. It could be argued that the presence of the "expansion switches" $s_i$ in Equation (27) represent additional free parameters in the model we have fit to the data. If this is so, the effective number of objects, $N_{tot} - p$, which enter into the variance computation should be correspondingly reduced. However, it is not entirely clear how many degrees of freedom these switches represent. It is clear that in at least a few cases (1559+19, Z7423), our use of the expansion switch may have been no more than a trick to make a few large-residual data points fit better, while in several other cases (Cancer, E508, OC3627) the data virtually require the expansion model. In the former case the use of the expansion model should be penalized in the scatter computation, while in the latter it arguably should not be. There is no "right" answer to this question, but a good approximation is to take the number of additional free parameters to equal the number of expanding clusters. If we assume that indeed there are nine additional degrees of freedom in our model, and calculate the scatter accordingly, we obtain the quantity $\sigma_s$ indicated in the fifth column of Table 3. This value, $\sigma = 0.398$ mag, represents our best estimate of the HM TF scatter. Evidently, the precise treatment of the expansion switches is not a major source of uncertainty.

### 3.2.2. Zeropoint Calculation

We have already discussed (§ 2.) the principles behind the HM zeropoint calculation; we now fill in the details. Our working hypothesis is that the HM cluster distances are on average equal to their radial velocities.[6] We express this average in logarithmic terms,

---

[6]We carry out this analysis in the CMB frame, although to the extent that the HM sample is truly isotropically distributed, it should hold in any reference frame moving uniformly with

since it is log(distances), rather than distances themselves, whose TF errors are normally distributed. A preliminary zeropoint calculation thus proceeds from the equation

$$\sum_{i=1}^{N} \log \left( \frac{v_i}{d_i} \right) = 0 \qquad (28)$$

where the sum runs over the 31 HM clusters. Recalling the definition of the arbitrary relative distance moduli $a_i$, we may write the cluster distances (in velocity units) as

$$d_i = 10^{0.2\mu_i} = 10^{0.2a_i} \times 10^{-0.2A}, \qquad (29)$$

where $A$ is the desired TF zeropoint. Using Equation (29) in Equation (28), we obtain

$$A = \frac{1}{N} \sum_{i=1}^{N} \left( a_i - 5 \log v_i \right). \qquad (30)$$

It is instructive to rewrite this last equation in terms of the cluster "hubble ratios"

$$h_i \equiv v_i 10^{-0.2a_i}, \qquad (31)$$

each a measure[7] of the expansion rate in units relevant to the TF observables:

$$A = -\frac{5}{N} \sum_{i=1}^{N} \log h_i. \qquad (32)$$

This formulation underlines the true significance of the TF zeropoint: it is a measure of the Hubble expansion in units suitable to converting the TF observables into distances in $\mathrm{km\,s^{-1}}$. However, expressing the zeropoint in this way suggests that the uniform weighting indicated by Equation (32) is not optimal. Errors affect any individual cluster hubble ratio in two ways. First, there is the "cosmic error" in the velocity $v_i$, i.e., the difference between $v_i$ and the Hubble velocity actually corresponding to the cluster distance. This error is, of course, nothing but the cluster's radial peculiar velocity $u_i$, but since we have no *a priori* knowledge of its value it enters into the calculation only as an rms error $\delta v$, whose specific value we consider below. Second, there is the random error in the determination of the

---

respect to the CMB.

[7]We note that our hubble ratio is similar to the "Hubble Modulus" of Rubin *et al.* (1976). The difference lies only in our adoption of $\mathrm{km\,s^{-1}}$ units for distance, whereas Rubin *et al.* measured distance in Mpc.

relative distance modulus $a_i$, which is given to good accuracy by $\delta a_i = \sigma/\sqrt{n_i}$ using simple $\sqrt{N}$ statistics.[8]

The effect of these two sources of error may be calculated as

$$\delta(\log h_i) = \delta\left(\log v_i - 0.2 a_i\right) = \frac{1}{5}\sqrt{f^2\left(\frac{\delta v}{v_i}\right)^2 + \frac{\sigma^2}{n_i}}, \qquad (33)$$

where $f \equiv \left(\frac{5}{\ln 10}\right)$ and in the last step we have assumed that the two sources of error are uncorrelated and therefore add in quadrature. Having determined (up to the as-yet-to-be-specified $\delta v$) the rms error in the $h_i$, we modify Equation (32) to read

$$A = -5\sum_{i=1}^{N} w_i^h \log h_i, \qquad (34)$$

where the weights $w_i^h$ are given by

$$w_i^h = \frac{(\delta\log h_i)^{-2}}{\sum_{i=1}^{N}(\delta\log h_i)^{-2}}, \qquad (35)$$

with the $\delta(\log h_i)$ given by Equation (33). Similarly, we compute the rms error of this TF zeropoint by adding in quadrature the errors in each term. This yields, after some algebraic manipulation,

$$\delta A = 5\left[\sum_{i=1}^{N}(\delta\log h_i)^{-2}\right]^{-\frac{1}{2}}. \qquad (36)$$

There remains only the question of the value of $\delta v$ to adopt in Equation (33). A safe lower limit is $\sim 100$ km s$^{-1}$, suitable if we believe that in general the Hubble flow is very "cold." Another possibility would be the (one-dimensional) velocity of the Local Group with respect to the CMB, $\sim 600/\sqrt{3} \simeq 350$ km s$^{-1}$. In practice, the exact choice of $\delta v$ has little effect, and we adopt $\delta v = 250$ km s$^{-1}$. When this value is substituted into Equation (33) and the resultant values of the $\delta h_i$ into Equations (34) – (36), we obtain the TF zeropoint and its rms error given in Table 3. If we adopt $\delta v = 100$ km s$^{-1}$, the zeropoint decreases by 0.005 mag; if we adopt $\delta v = 350$ km s$^{-1}$, the zeropoint increases by 0.002 mag. These changes are clearly negligible, so we need not be concerned about the the precise value of $\delta v$ for the zeropoint determination.

---

[8]This expression is not exact, owing to covariance between the the fitted value of the TF slope $b$ and each of the $a_i$. However, since many clusters go into determining a single slope, this covariance is negligible.

By contrast, the precise value of $\delta v$ has a rather more significant effect on the calculation of the zeropoint *uncertainty*, $\delta A$. This error is reduced to $\sim 0.02$ mag for $\delta v = 100 \text{ km s}^{-1}$, and is increased to 0.035 mag for $\delta v = 350 \text{ km s}^{-1}$. Perhaps more importantly, we have not quantified the uncertainty in the zeropoint arising from possible errors in our fundamental assumption, namely, that the HM clusters depart systematically from uniform Hubble expansion by at most a bulk flow. It is easy to imagine a variety of patterns of large-scale departure from Hubble flow which would vitiate this assumption, and thus cast our zeropoint calculation into doubt. As it is not within the scope of this paper to assess the nature or likelihood of such flows, it is pointless to try to quantify further the probable error in the TF zeropoint. In a later paper in this series (Dekel *et al.* 1995), we will discuss how the POTENT analysis can constrain possible global zeropoint errors in the TF calibration.

Finally, having assigned a zeropoint to the HM TF relation, we may display the relation for the full 346 galaxy, 31 cluster sample. This is done in the left hand panel of Figure 11, where we plot absolute magnitude, computed as $m^{(c)} - \mu$, vs. circular velocity parameter. The $m^{(c)}$ are the bias-corrected absolute magnitudes following convergence; the distance moduli are those assigned by the model, including cluster expansion when appropriate. In the right hand panels we plot residuals from this TF relation vs. $\eta$ for the HM North and South sample separately. No meaningful trend of the residuals with $\eta$ is evident in either plot, indicating that the linear TF relation adopted here is a good description of the data.

### 3.2.3. *Determination of the Internal Extinction Coefficient*

We have, until now, left undetermined the value of the coefficient $C_{\text{int}}^{\text{I}}$ used in correcting the HM apparent magnitudes. This coefficient (see § 2) multiplies log(axial ratio) ($\mathcal{R}$) to give the I band internal extinction correction. In fact, the TF calibration discussed above was carried out using $C_{\text{int}}^{\text{I}} = 0.95$. We arrived at this value by considering the variation in the TF scatter with respect to $C_{\text{int}}^{\text{I}}$, which is plotted in the left-hand panel Figure 12. The $\sigma$'s shown in the figure were obtained from the full calibration procedure described in above, including cluster expansion and bias correction, for each value of $C_{\text{int}}^{\text{I}}$ (however, they do not assume the expansion switches add degrees of freedom, which is only a uniform scaling in any case). We see that the TF scatter indeed depends on $C_{\text{int}}^{\text{I}}$, and that it has a fairly well-defined minimum around $C_{\text{int}}^{\text{I}} \simeq 0.9$–$1.0$. However, the value of $C_{\text{int}}^{\text{I}} \simeq 0.95$ is not necessarily highly accurate. To estimate error bars on $C_{\text{int}}^{\text{I}}$, we may write an approximate $\chi^2$ for the fit as $N_{\text{eff}}\sigma^2$, where $N_{\text{eff}}$ is the effective number of independent data points, which is $\sim 310$ in the present case (depending on the treatment of the clusters allowed to expand). We obtain a 65% confidence limit by asking for what values of $C_{\text{int}}^{\text{I}}$ $\chi^2$ changes by 1 unit. Such a change in $\chi^2$ corresponds to a change in the TF scatter of 0.0041 mag. This scatter increase over the minimum is shown as a dotted line in Figure 12. An approximate interpolation of the curve through the dotted line then shows that with 65% confidence we may place the I band internal extinction coefficient in the range $0.65 \lesssim C_{int}^{I} \lesssim 1.3$. The 90% confidence limits on $C_{int}^{I}$ are, fortunately, only somewhat wider, roughly $0.4 \lesssim C_{int}^{I} \lesssim 1.6$.

In the right hand panels of Figure 12, we plot residuals from the TF fit, defined as the observed (bias-corrected) minus the expected apparent magnitude, for three different values of the internal extinction coefficient: $C_{int}^{I} = 0.0$ (top panel), $C_{int}^{I} = 2.0$ (middle panel), and $C_{int}^{I} = 0.95$ (bottom panel). The top panel convinces us of the reality of the internal extinction effect: when no correction is made, highly inclined galaxies are faint and less inclined galaxies are bright, relative to their Tully-Fisher prediction. The middle panel demonstrates that adopting too large a value of $C_{int}^{I}$ clearly overcompensates. The bottom panel shows that with $C_{int}^{I} = 0.95$, the TF residuals show no meaningful trend with axial ratio. We are, in fact, confident that this value is more accurate than we have estimated from the HM sample alone. As we shall see in Paper II, the much larger Mathewson sample yields a nearly identical value; in addition, Giovanelli *et al.* (1994) have recently obtained a similar result ($C_{int}^{I} = 1.05$). We note, finally, a peculiarity of the HM sample evinced in the right hand panels of Figure 12. There is a cutoff in the axial ratio at $\log(a/b) \simeq 0.7$. This apparent cutoff is an artifact that occurs because Han, Mould and collaborators did not tabulate axial ratios but only estimated inclinations. Since the formula giving inclination as a function of axial ratio assigns an inclination of 90° above $\log(a/b) = 0.7$, there is no way to determine the original axial ratios for these objects. The fact that we were able clearly to identify the internal extinction effect from this sample, despite the artificial axial ratio cutoff, testifies to its significance.

## 3.3. Calibration of the Inverse HM TF relation

We are motivated to calibrate the inverse TF relation in the hope of achieving a bias-free approach to peculiar velocity analysis (Dekel 1994). While it is often a good approximation to treat the inverse relation as free of selection bias, such bias may not always be entirely negligible. The key issue is whether or not the sample selection criteria are $\eta$-independent. In fact, the derived relations between UGC diameters or Zwicky magnitudes and the TF observables do exhibit an explicit $\eta$-dependence (see Table 2). In addition, the H I flux-per-channel quantity on which the HM South sample apparently depends is a function of $\eta$ (albeit prior to inclination correction). However, we saw that the ESO diameter showed no explicit dependence on $\eta$, but instead a strong redshift dependence. This raises the question as to whether the $\eta$-dependence of the UGC diameter is entirely real. In addition, while it was apparent that the H I flux-per-channel played some role in the selection of the HM South subsample, the precise limiting value remained somewhat vague (§ 3.1.2). For these reasons, the $\eta$-dependent character of the HM sample selection procedure is rather uncertain. We therefore present two calibrations of the inverse TF for HM: one with and one without bias corrections applied. We shall see that the differences between the two calibrations are small but not entirely negligible.

### 3.3.1. Method of Fit

We adopt the 9-cluster expansion model in carrying out the inverse fit. Given that model, the expected circular velocity parameter of the $j^{th}$ galaxy in the $i^{th}$ cluster is given by (the notation is that of § 2.1)

$$E(\eta_{ij}|m_{ij}) = -e\,(m_{ij} - \mu_{ij} - D)\;,\tag{37}$$

with

$$\mu_{ij} = \mu_i + s_i \times 5\log\left(\frac{v_{ij}}{v_i}\right)\;,\tag{38}$$

where as before $s_i = 1$ in the case of the 9 "expanding" clusters, $s_i = 0$ otherwise. We again introduce arbitrary relative distance moduli $a_i$, now defined by

$$a_i = D + \mu_i\;.\tag{39}$$

Although we use the same symbol as we did for the forward relation, the numerical values of the $a_i$ will in general differ from the earlier case.

Equations (37)–(39) lead to the following sum-of-squares quantity:

$$\mathcal{S} = \sum_{i=1}^{N}\sum_{j=1}^{n_i} w_i\left[\eta_{ij} - \left(-e\left[m_{ij} - a_i - s_i \times 5\log\left(\frac{v_{ij}}{v_i}\right)\right]\right)\right]^2\tag{40}$$

where the $w_i$ are given as before by Equation (19). Minimization of $\mathcal{S}$ with respect to the variations in $e$ and the $a_i$ yields the best-fitting values of these parameters. The inverse TF scatter $\sigma_\eta$ is estimated as the rms $\eta$-residual with respect to the fit, taking into account the effective number of degrees of freedom as discussed above. We calculate the inverse TF zeropoint in exact analogy with the forward case, i.e.,

$$D = -5\sum_{i=1}^{N} w_i^h\,(\log v_i - 0.2a_i)\;,\tag{41}$$

where the weights $w_i^h$ are given by equations (33) and (35) above. (In the present case, the scatter $\sigma$ in Equation (33) is given by $\sigma_\eta/e$.) When this procedure is carried out (neglecting selection bias) we obtain the inverse TF calibration presented in the first line of Table 4.

If we choose not to neglect the $\eta$-dependence of the HM sample selection criteria, then we must correct the inverse TF calibration for selection bias. As discussed by W94, this bias correction is perfectly analogous to that applied to the forward TF calibration. However, for the inverse TF it is the circular velocity parameters, not the apparent magnitudes, which are modified in successive iterations of the correction scheme. That is, at each iteration we determine the relative bias $\mathcal{B}$ and compute "corrected" circular velocity parameters

$$\eta^{(c)} = \eta + \sigma_\eta \mathcal{B} . \tag{42}$$

We then refit the model by replacing $\eta_{ij}$ with $\eta_{ij}^{(c)}$ in Equation (40), and iterate. The relative bias $\mathcal{B}$ for the inverse relation is identical in mathematical form to the corresponding quantity in the forward case, but is much smaller in absolute size, reflecting the weak dependence of sample selection on $\eta$ (W94). The inverse scatter $\sigma_\eta$ is also corrected for bias. When we apply this procedure to minimization of Equation (40), we obtain the inverse TF calibration given in the second line of Table 4. The inverse TF relation is *not* the mathematical inverse of the forward TF, *i.e.*, $D \neq A$, and $e^{-1} \neq b$, for either of the lines in Table 4. This is to be expected by analogy to simple least-square fits: a fit of $y$ on $x$ is not the inverse of $x$ on $y$ for data with scatter (see, for example, Willick 1991, Appendix 3, for a detailed discussion). However, the effective scatter in magnitudes of the inverse TF, $\sigma_\eta/e$, is quite similar to the foward TF $\sigma$.

## 4.    The W91CL Sample

The full W91CL sample consists of 156 galaxies in 11 clusters. One galaxy, UGC 12382 in Pegasus, was excluded because of its very large ($> 3\sigma$) residual with respect to any model (this object was excluded for the same reason from the HM TF analysis). Of the 11 clusters, 10 coincide with the HM92 North clusters (see Table 1). These ten clusters formed the basis of the H band TF study of Aaronson *et al.* (1986), and the raw 21 cm velocity widths used by W91CL (and HM92 North) derive from that earlier study. The eleventh W91CL cluster consists of 25 galaxies in nearby ($cz \sim 1000$ km s$^{-1}$) Ursa Major, for which the raw velocity widths are obtained mainly from Aaronson *et al.* (1982) but also partly from Pierce and Tully (1988). All raw velocity widths, in W91CL as in HM, are on the $\Delta V_{20}^{(c)}$ system used by the Aaronson group. The $\eta$-values of W91CL galaxies are not equal, in general, to those used for the same galaxies in these earlier studies (or to those used for the same HM92 North galaxies), since the axial ratios and thus the inclination corrections have been redetermined. The W91 apparent magnitudes are obtained from Lick r band CCD imaging photometry. We describe this bandpass in Paper III; we note for the moment only that it is quite similar but not identical to the better known Gunn r bandpass (Thuan and Gunn 1976).

## 4.1.    Selection Criteria Relations for W91

The ten "distant" clusters in W91CL common to HM92 North may be expected to share the latter's selection criteria (§ 3.1.1). We thus take these ten W91CL clusters to be selected to the limits of the Zwicky ($m_z = 15.7$ mag) and UGC ($D_{UGC} = 1.0'$) catalogs. The Ursa Major galaxies are a much brighter sample, but are so close that calibration bias is negligible. The W91CL TF calibration will therefore use the "two-catalog" bias correction formulae of W94 for objects in the ten distant clusters, while for the 25 Ursa Major galaxies we neglect bias correction altogether. We assume, as we did in the case of HM North, that any incompleteness associated with H I nondetection is negligible.

As in the case of HM we begin by determining for the W91CL samples the coefficients of the relations

$$\log D_{UGC} = a_1 - b_1 m_r - c_1 \eta + d_1 \mathcal{R} - e_1 A_B \tag{43}$$

and

$$m_z = a_2 + b_2 m_r + c_2 \eta + d_2 \mathcal{R} + e_2 A_B , \tag{44}$$

through multiparameter fits using the photographic, CCD, and kinematic data. Determination of the coefficients is again carried out in an iterative fashion, correcting for the biases due to the limits on the UGC diameters and Zwicky magnitudes. The results are given in Table 5. We have included the W91PP (§ 1) galaxies in these fits, since this greatly adds to the statistical accuracy of the fitted coefficients. In the case of the UGC diameter fit, the addition of the W91PP galaxies presents no difficulty, as that sample is also complete to the UGC diameter limit (Paper II). However, in the case of the Zwicky magnitude fit, a potential bias arises since W91PP is *not* complete to the Zwicky catalog limit. What we have done in practice is to use the full W91 sample to better constrain the less well determined coefficients in Equation (44) ($c_2$, $d_2$, and $e_2$), but the W91CL sample alone to calculate the well-constrained parameters ($a_2$ and $b_2$).

The results are qualitatively similar to those obtained from the HM North sample (first two lines in Table 2). This is not surprising, as the only important difference is that the W91 uses r band CCD photometry and HM I band CCD photometry. In neither the HM nor the W91 data do we see a dependence of $\log D_{UGC}$ on $A_B$. In addition, the scatter in these relations is quite similar. However, whereas we saw no $A_B$-dependence in the $m_z - m_I$ relation for HM, that dependence is seen clearly for the W91 sample, perhaps because of the larger number of objects involved in and smaller scatter of the $m_z - m_r$ fit. In Figure 13, we plot the bias-corrected values of $\log D_{UGC}$ (top panel) and $m_z$ bottom panel against the raw r band magnitudes. Note that (as was the case with HM) the bias-corrected values can exceed the formal limits and fainter values of $m_r$; as before, a consequence will be selection bias in the TF calibration.

### 4.2. Calibration of the W91 TF Relation Using W91CL

We use the W91CL subsample to calibrate the W91 TF relation; in Paper II, we will show that this calibration is consistent with one obtained independently from the W91PP subsample. The calibration procedure for W91CL is analogous to that applied to HM. We assume first that W91CL consists of eleven true clusters, each composed of galaxies at a single distance. With this preliminary model, we fit a TF relation by minimizing a sum-of-squares analogous to that in Equation (18) and applying the bias-correction procedure of W94. We also compute the W91 zeropoint using the prescription adopted for the HM clusters (§ 3.2.2). We thus obtain a TF relation whose parameters are given in the first line of Table 6.

Since ten of the eleven W91 clusters are in common with the HM sample, it is not surprising that we find a trend of TF residuals with redshift in several of the clusters. The residuals for these clusters are shown in Figure 14. It is quite clear that in the Ursa Major, Pegasus, and Cancer clusters, the residuals correlate strongly with radial velocity, indicating that these clusters are better modeled as "expanding" in the sense described in § 3.2.1. Three other clusters, Coma, A2634, and A2151 show a significant hint of expansion. When we allow these clusters to expand, using the model developed in § 3.2.1, and repeat the calibration analysis, we obtain the TF parameters given in the second line of Table 6. The addition of six expansion switches has reduced the scatter significantly, from $\sim 0.41$ mag to $\sim 0.35$ mag. This may be compared with a statistical uncertainty in the scatter of $\sim 0.35/\sqrt{275} \simeq 0.021$ mag. A plot of TF residuals against log redshift for the same clusters when the expansion model is used is shown in Figure 15. The trends seen in the previous figure have now been largely eliminated.

Note that while HM92 North TF calibration also suggested that the Pegasus, Cancer and Coma clusters are expanding, that trend was not visible in the HM sample for A2634 and A2151. Conversely, the HM data suggested that Z7423 and A1367 are expanding, while the W91CL data do not. These discrepancies are not serious; they merely reflect what we have already stated in § 3.2.1, namely, that we have allowed ourselves some extra freedom in deciding which clusters can be modeled as expanding. We must, as a result, pay the price by reducing the numbers of degrees of freedom in the model accordingly. This has been done in the statistic $\sigma_s$ in Table 6. The value of $\sigma_s$ in the second line of the table represents our best estimate of the W91 TF scatter. In the third line of Table 6, we show the parameters resulting from fitting the calibration model without applying the iterative bias correction procedure. As in the case of the HM calibration, we see that the uncorrected TF relation has a flatter slope and a smaller scatter than the fully corrected one. The zeropoint differs significantly as well, as we discuss further in § 5.

### 4.2.1. Determination of the W91 Internal Extinction Coefficient

As above, we determine the internal extinction coefficient for W91 by minimizing the TF scatter with respect to variations in $C_{int}^r$. The W91 TF scatter–$C_{int}^r$ dependence is shown in the left hand panel of Figure 16. A reasonably well-defined minimum exists, centered on $C_{int}^r \simeq 1.2$. The three right hand side panels of the figure demonstrate the behavior of residuals for no correction (top), overcorrection (middle), and proper correction (bottom panel). We have adopted a final value $C_{int}^r = 1.15$, somewhat smaller than the absolute minimum in the left hand panel ($C_{int}^r = 1.2$), in anticipation of a similar graph we will show in Paper II for the W91PP galaxies. The W91PP sample prefers a slightly smaller value, $C_{int}^r \sim 1.0$, than does the cluster sample; $C_{int}^r = 1.15$ represents a compromise weighted slightly toward W91CL. As was the case with HM, the data do not in fact constrain the value of $C_{int}^r$ very tightly around the minimum. Using a statistical argument similar to that given in § 3.2.3, we find that $\delta\chi^2 = 1$ corresponds to a scatter increase of 0.01 mag, shown as a dotted line in the figure. Interpolating the curve through this line, we can state with 65% confidence that $C_{int}^r$ lies in the range $\sim 0.65$–1.75. Consideration of the W91PP data will narrow this range considerably and will confirm that our adopted value of $C_{int}^r = 1.15$ adequately describes the data.

### 4.2.2. Calibration of the Inverse W91 TF relation

We calibrate the W91 inverse TF relation in complete analogy with HM (§ 3.3.1). We retain the 6-cluster expansion model which signficantly reduces the scatter of the W91 forward TF relation. Table 7 gives the parameters of the W91 inverse TF relation obtained with and without selection bias corrections. The slope and scatter of the W91 inverse TF relation are quite insensitive to whether or not the bias-correction procedure is applied. The zeropoint is somewhat more sensitive to the correction but still changes far less than its forward counterpart. The changes in all the parameters are in any case no larger than the statistical uncertainties characterizing the fit.

## 5. Validity of the Bias Corrections and Comparison of HM with W91CL

The procedure we have used to correct for selection bias has a substantial effect on the Mark III catalog distances. One can appreciate the significance of the bias-correction procedure by comparing the TF parameters derived with and without it. Table 3 shows that the HM zeropoint increases by 0.24 mag, and the slope increases by $\sim 15\%$, as a result of these corrections. The sense of these changes is that the corrected TF relation yields inferred distances which are $\sim 22\%$ shorter for intrinsically faint ($\eta \simeq -0.2$) galaxies, $\sim 11\%$ shorter for typical ($\eta \simeq 0$) galaxies, and roughly unchanged for the most luminous $\eta \gtrsim 0.2$ spirals. Clearly, such changes will have a strong impact on the peculiar velocity field derived from the data. The bias corrections have also markedly increased the estimated scatter of the relation. Similar though somewhat smaller changes in the W91CL TF parameters are evident in Table 6.

One can better understand the effect of the bias-correction procedure by considering the corrections made to each object in the calibration analysis. In Figure 17, we plot the relative bias $\mathcal{B}$ for HM galaxies as a function of predicted raw I band magnitude (top panel) and log redshift (bottom panel). Figure 18 shows the corresponding plots for W91CL. Several aspects of these figures deserve mention. First, the relative bias is $\lesssim 1$ for a large majority of objects. This must be the case in order for the correction procedure to work, as noted by W94. Still, the relative bias is often of order a few tenths or greater, confirming the impression of its significance that we gained from the change in the TF parameters. Second, the mean bias value changes only slowly with redshift. This is because the more nearby samples are sufficiently complete as to contain objects as apparently faint as the more distant samples. Third, the relative biases computed in the W91 calibration are smaller, in the mean, particularly for $\log(cz) \gtrsim 3.8$. There are two reasons for this. First, the W91 TF scatter is $\sim 10\%$ smaller than the HM TF scatter, and the relative bias is roughly proportional to the TF scatter (W94). Secondly, the $\log D_{\mathrm{UGC}}$–$m_r$ relation places fewer W91 objects much fainter than the UGC catalog limit than does the $\log D_{\mathrm{UGC}}$–$m_I$ relation for HM. We discuss this further in § 5.2.

## 5.1. Validating the Bias Corrections: Comparison of Forward and Inverse Fits

In view of the large effect the bias corrections have on the TF calibration, it is important to validate them. There is no way to guarantee that our results are free of all biases. However, we do have a means of demonstrating, first, that a naive forward TF calibration is subject to an easily visible bias, and second, that our correction procedure eliminates this bias. The test involves a comparison of forward and inverse TF distances of the sample clusters. Any bias present in the inverse TF calibration is small, as evidenced by the relatively small changes in the inverse TF parameters when the corrections are applied (Tables 4 and 7). The inverse TF calibrations (with or without bias corrections) thus serve as benchmarks relative to which the effects of bias correction on the forward calibration may be assessed.

In Figure 19 we plot the difference between the forward and inverse cluster distance moduli as a function of log redshift for the HM clusters. In the upper panel moduli obtained from the *uncorrected* fits are shown; in the lower panel the bias-corrected moduli are plotted. The upper panel evinces a clear trend: the forward distance moduli of the lower-redshift clusters are systematically too large, and those of higher-redshift clusters systematically too small, relative to the inverse distance moduli. A linear fit of $[\mu(\text{forw}) - \mu(\text{inv})]$ to $\log(cz)$ yields a slope of $-0.46 \pm 0.08$, indicating that the trend is highly significant. There are two related causes of this trend. First, the mean bias per cluster is somewhat larger (Figure 17) at higher redshifts. The bias causes galaxies to appear brighter than their TF-expected apparent magnitude; the clusters thus appear correspondingly too close. Second, the too-flat slope of the uncorrected fit, itself a consequence of selection bias, causes large-$\eta$ (typically higher-redshift) objects to appear closer, and small-$\eta$ (typically lower-redshift) objects farther, than they actually are (over and above the apparent magnitude biases for these objects). To the extent that the inverse distances are also biased due to $\eta$-dependence of sample selection, the trend manifested in the figure is not as strong as it would be if we could plot the difference between uncorrected forward and *true* distance moduli. Thus, the bias we correct for is both real and signficicant.

In the lower panel of Figure 19 we see that the downward sloping trend has vanished. A linear fit of $\mu(\text{forw}) - \mu(\text{inv})$ to $\log(cz)$ now yields a slope of $0.02 \pm 0.09$, confirming the absence of any trend with redshift. It is gratifying to note that, despite their very different sample selection criteria, the HM South and HM North clusters show no meaningful offset from one another in the graph.[9] We have taken the error bars in Figure 19 to be $\propto \sigma/\sqrt{N}$, with the proportionality constant determined by requiring the $\chi^2$'s of the aforementioned linear fits to be roughly equal to the number of degrees of freedom $(31 - 2 = 29)$ for those fits. It turns out that the coefficient required is 0.6, indicating that the forward and inverse distance moduli exhibit errors which are substantially correlated. This is encouraging, as a major flaw in the bias correction procedure would result in a smaller degree of forward-inverse correlation.

Figure 20 shows the same plot for W91CL. Again, a strong trend is evident in the upper panel of the plot, in which bias corrections have been neglected. The trend does not extrapolate linearly to the lowest redshift cluster, Ursa Major, which is so close that bias corrections are negligible even for the forward fit. Its uncorrected forward distance is still significantly biased, because the selection bias affecting the distant clusters produces an erroneous TF zeropoint. Once the bias corrections are applied (lower panel), the trend with redshift vanishes; in particular, the Ursa Major forward and inverse distances are now fully consistent. Thus, for the W91 as for the HM TF calibration, the correction procedure adequately remedies the principal effects of selection bias.

---

[9]As was noted in §3.1, we have selected the HM South effective diameter limits in part to enforce this outcome. This was a necessary step, prompted by the absence of strict selection criteria for the HM South.

## 5.2. Comparison of the HM and W91 Calibrations

Our calibration of the W91 TF relation is only partial, as the W91CL sample is not distributed over the sky well enough for a proper zeropoint determination. Our calibration of the HM TF relation is complete, however. By comparing the two calibrations for objects in common, we can illustrate any zeropoint discrepancy which remains between the two samples. Such a comparison can also shed further light on the bias corrections.

In Figure 21, we plot distance modulus differences for the ten clusters in common between the two samples. The upper panel shows the moduli derived from the forward fits with full bias corrections, and the bottom panel those from the inverse fits without bias corrections. Both panels show a clear offset from zero, $i.e.$, the HM modulus is greater than the W91 modulus in the mean. For the forward fit we find $< \mu(HM) - \mu(W91) >= 0.080 \pm 0.028$ mag; for the inverse fit $< \mu(HM) - \mu(W91) >= 0.109 \pm 0.028$ mag.[10] Apparently the HM calibration has resulted in distances which are typically $\sim 4\%$ larger than those derived from the W91 calibration.

The fact that the distance modulus offset occurs in both panels of Figure 21 shows that it has nothing whatsoever to do with the bias correction procedure; the inverse fits have undergone no such corrections. The offset instead reflects the fact that, while the ten clusters shown have fully determined a provisional W91 zeropoint, the HM zeropoint is determined from 31 clusters distributed around the sky. To confirm this, one can repeat the calculation of the HM zeropoint (§3.3.2) but sum over only the 10 clusters in common. When this exercise is carried out, the HM forward TF zeropoint increases by almost exactly 0.08 mag, precisely cancelling the HM/W91 distance modulus difference. The inverse TF zeropoint similarly changes almost precisely by the amount needed for distance agreement. Thus, the ten clusters common to HM and W91CL alone yield TF zeropoints leading to excellent distance agreement in the mean (although the agreement cluster by cluster is not perfect, as is indicated by the scatter in Figure 21). But these ten clusters alone do not produce *correct* distances, since they do not measure the same expansion rate as the full HM cluster sample. This fact underscores the necessity of using a full-sky sample for determining the proper TF zeropoint, a point we have emphasized repeatedly.

---

[10]The error bars shown are those for the W91 clusters only, $i.e.$, the errors for the two samples are not added in quadrature. The $\chi^2$ computed from these error bars is $\sim 10$, indicating that the W91 errors correctly measure the typical $difference$ between W91 and HM.

A final comparison between HM and W91 may be carried out on a galaxy-by-galaxy, rather than a cluster-by-cluster basis. In Paper II we will use galaxy-by-galaxy intercomparisons to determine the final zeropoint for W91, as well as for the remaining samples. Such a procedure is applied to raw (as opposed to bias-corrected) distance moduli, and so differs fundamentally from the cluster-cluster comparison shown above. Still, the two approaches must yield consistent results if the entire calibration procedure is to be trustworthy. We compute individual galaxy distance moduli as $\mu = m - (A - b\eta)$. In Figure 22 we plot individual galaxy distance modulus differences vs. log redshift. The heavy line indicates the mean value $<\mu(HM) - \mu(W91)> = 0.05 \pm .02$ mag for 114 galaxies in common between W91CL and HM; the dotted line shows the mean cluster difference of 0.08 mag. Thus, the individual object distance moduli show a smaller offset, by 0.03 mag, than do the cluster distance moduli. In addition, the individual object modulus differences have a very small scatter, 0.20 mag, whereas the cluster modulus differences correspond (see above) nearly to the full scatter (0.35 mag) of the W91 TF relation.

There are two sources of these discrepancies. The first is that the makeup of any given cluster sample is not the same in W91CL and HM. The second is additional error introduced by the bias correction procedure. To assess the latter, we must first estimate the former. To do so, we carry out the object-object comparison using the inverse TF calibrations without bias corrections. Defining individual object inverse distance moduli as $\mu = m - (D - \eta/e)$, we find $<\mu(HM) - \mu(W91)> = 0.10 \pm .02$ mag, *i.e.*, 0.01 mag less than the result from the cluster-cluster comparison using the inverse TF. We take this as a measure of the difference between the two kinds of comparisons caused by the fact that the HM and W91CL cluster samples do not contain exactly the same galaxies. It then follows that $\sim 0.03 - 0.01 = 0.02$ mag of the discrepancy between the forward TF cluster-cluster and object-object comparisons results specifically from inconsistency in the bias-correction procedure as applied to HM vs. W91CL. The enhanced scatter of the cluster-cluster comparison relative to the galaxy-galaxy comparison is also a consequence of this inconsistency.

While the 0.02 mag discrepancy represents a cautionary note in our overall approach, it is not a cause for great concern, amounting to only $\sim 1\%$ in distance. Indeed, we might view it as something of a triumph that the agreement is as good as it is. We have seen that the amplitude of the bias corrections for HM and W91 calibrations are noticeably different (Figures 17 and 18). The full bias-correction procedure, including characterization of sample selection and the fitting of the $(\log D_{\mathrm{UGC}}, m_z)$–$(m, \eta)$ relations is quite involved. Uncertainty in these latter relations—which stems from slight mischaracterizations of sample selection criteria, as are inevitable here—feeds back into the TF calibration bias corrections. In view of all this, the small difference between the cluster-cluster and object-object distance modulus comparisons is indicative of a reasonable degree of robustness in the full calibration procedure. Residual errors of $\sim 0.02$ mag are irreducible elements of our forward TF calibrations, traceable to the inexactness of the selection criteria for most samples.

## 6. Summary

The principal result of this paper has been the calibration of the Tully-Fisher relation for the I band cluster sample of Han and Mould (HM). The parameters which specify this calibration are given in second line of Table 3. We have, in addition, provided a partial calibration (slope and scatter only) of the TF relation for the r band cluster sample of Willick (W91CL) (Table 6). We defer a final determination of the W91 TF zeropoint to Paper II, in which we also treat the field galaxy spiral samples of Willick (W91PP), Courteau-Faber (CF), Mathewson et al. (MAT), and Aaronson et al. (A82). In Paper III, we will present fully corrected Tully-Fisher distances, along with a variety of ancillary data, for all five spiral samples. Along with the previously published elliptical galaxy data of Lucey and Carter (1988), Faber et al. (1989), and Dressler and Faber (1990), these data constitute the Mark III Catalog of Galaxy Peculiar Velocities. In later papers in this series, we will apply the POTENT algorithm to the Mark III data.

We have carried out our analysis in a system of units in which distances $r$ are measured in $\mathrm{km\,s^{-1}}$ and the distance modulus $\mu$ is defined simply as $\mu = 5 \log r$. With these conventions, conversions between the TF observables $(m, \eta)$ and distances in $\mathrm{km\,s^{-1}}$ may be made straightforwardly, without reference to either a Hubble constant or to any arbitrary "reference" distance, as has been used in some previous work. We have avoided the use of models of the peculiar velocity field in our TF calibration procedure. However, in order to obtain a zeropoint for the HM TF relation, it was necessary to assume that the radial peculiar velocities of the HM clusters vanish when averaged over the entire sample. To the extent that the HM sample is uniformly distributed on the sky, this assumption does not exclude the possibility of a large-scale bulk motion with respect to the CMB frame. However, it does require that the "local" value of the Hubble constant—the expansion rate of the volume which the HM clusters occupy—is not significantly different from its global value.

A central element in the TF calibration procedure was correction for selection (or "calibration") bias. This effect, when neglected, results in apparent TF relations which are flatter and have smaller scatter than the true relations. In addition, relative distances are compressed, resulting in a TF zeropoint which is biased low. We have corrected for selection bias by implementing the iterative correction procedure of W94. This procedure requires that the selection criteria for each sample be characterized accurately and quantitatively. We have endeavored to do this but have emphasized that our characterization of the sample selection criteria is in some cases only approximate. This is unfortunate but inevitable, due to the nonrigorous nature of the construction of current TF samples. Despite the approximations we have made in describing sample selection, we have confirmed the essential validity of the bias-correction procedure through a comparison of cluster distance moduli derived from the forward and (nearly unbiased) inverse forms of the TF relation.

We have standardized the transformations that the raw observables (velocity width and apparent magnitude) undergo prior to their use in the Tully-Fisher analysis. We will describe the details of these transformations in Paper III. One aspect has, however, been dealt with here, namely, the determination of the internal extinction coefficient $C_{\rm int}$. We have taken the view that this coefficient is best estimated by adjusting it to minimize the scatter in the TF relation. Doing so, we have found from the HM sample that $C_{\rm int}^{\rm I} = 0.95$, and from the W91CL sample that $C_{\rm int}^{\rm r} = 1.15$. The uncertainty in these values was large, but we will present further evidence that they are essentially correct in Paper II when we analyze the MAT and W91PP samples. Since the value of $C_{\rm int}$ has a significant effect on the sample apparent magnitudes, it should be viewed as an integral part of the TF calibration.

We would like to thank Michael Strauss for a careful reading of an early version of this paper. We are grateful to Ming-Sheng Han and Jeremy Mould for providing us with the raw data from their cluster TF data base. This work has been partially supported by the US-Israel Binational Science Foundation.

Clusters Comprising the HM Sample

| Cluster Name | $l(°)$ | $b(°)$ | $v_{\mathrm{CMB}}\,(\mathrm{km\,s^{-1}})$ | $N_{\mathrm{HM}}$ | $N_{\mathrm{W91}}$ |
|---|---|---|---|---|---|
| HM92 North: | | | | | |
| PISCES | 126 | -33 | 4723 | 20 | 19 |
| A400 | 170 | -45 | 7500 | 7 | 7 |
| A539 | 196 | -18 | 8532 | 10 | 9 |
| CANCER | 203 | 29 | 4788 | 20 | 18 |
| A1367 | 235 | 73 | 6976 | 19 | 19 |
| COMA | 57 | 88 | 7299 | 13 | 12 |
| Z74-23 | 350 | 66 | 6140 | 7 | 6 |
| A2151 | 32 | 45 | 10718 | 9 | 10 |
| PEGASUS | 88 | 48 | 3812 | 13 | 20 |
| A2634 | 104 | -36 | 8119 | 10 | 10 |
| HM92 South: | | | | | |
| ANTLIA | 273 | 20 | 3236 | 11 | |
| CEN30 | 303 | 22 | 3705 | 10 | |
| CEN45 | 303 | 22 | 5096 | 6 | |
| E508 | 309 | 39 | 3272 | 13 | |
| HYDRA | 297 | 27 | 4112 | 10 | |
| N3557 | 282 | 22 | 3400 | 6 | |
| M93: | | | | | |
| A779 | 191 | 45 | 7057 | 6 | |
| MKW1S | 232 | 32 | 5231 | 17 | |
| MKW4 | 271 | 63 | 6569 | 7 | |
| AWM2 | 230 | 80 | 7078 | 15 | |
| MKW11 | 334 | 75 | 7265 | 16 | |
| AWM3 | 26 | 70 | 4681 | 27 | |
| 3C296 | 346 | 64 | 7559 | 14 | |
| 1559+19 | 38 | 46 | 4722 | 8 | |
| M91: | | | | | |
| N3256 | 279 | 11 | 3067 | 9 | |
| OC3560 | 311 | 30 | 2928 | 13 | |
| N5419 | 321 | 28 | 4345 | 4 | |
| OC3627 | 322 | -6 | 4497 | 9 | |
| PAVOII | 328 | -23 | 4417 | 8 | |
| TEL | 354 | -38 | 2629 | 5 | |
| OC3742 | 353 | -41 | 4793 | 4 | |

Table 1: The clusters which make up the Han-Mould Tully-Fisher calibration sample.

Fit Coefficients for HM

| Quantity | | Coefficient of | | | | | |
|---|---|---|---|---|---|---|---|
| Predicted | const. | $m_I$ | $\eta$ | $A_B$ | $\mathcal{R}$ | $\log(cz) - 3.5$ | $\sigma$ |
| $\log D$(UGC) | 3.707 | 0.213 | 0.185 | 0.000 | 0.413 | 0.000 | 0.124 |
| $m_z$(Zwicky) | 0.306 | 1.110 | 2.267 | 0.000 | 0.718 | 0.000 | 0.447 |
| $\log D$(ESO) | 2.971 | 0.142 | 0.000 | 0.000 | 0.350 | 0.442 | 0.112 |
| $\log(F_{\mathrm{H\,I}})^a$ | 1.015 | 0.175 | 0.000 | 0.000 | 0.000 | 0.665 | 0.255 |

Table 2: Coefficients of the indicated quantities in the linear relations for $\log D_{\mathrm{UGC}}$, $m_z$, $\log D_{\mathrm{ESO}}$, and $\log(F_{\mathrm{H\,I}})$ in terms of the TF observables for the HM sample. Also given are the rms dispersions ($\sigma$) for these relations. Notes: (a) the quantity fitted was the fully corrected, rather than the raw, I band magnitude.

Parameters of the HM TF Relation

| | $A\,(\pm)$ | $b\,(\pm)$ | $\sigma$ (mag) | $\sigma_s$ (mag) | $N_\sigma$ |
|---|---|---|---|---|---|
| No Expand | $-5.415\,(.032)$ | $7.917\,(.187)$ | 0.456 | 0.456 | 346 |
| Expand 9 | $-5.481\,(.029)$ | $7.865\,(.162)$ | 0.392 | 0.398 | 346 |
| No Bias Corr | $-5.718\,(.028)$ | $6.925\,(.158)$ | 0.360 | 0.365 | 346 |

Table 3: Parameters resulting from fitting a forward Tully-Fisher relation to the HMCL sample. The 1-$\sigma$ uncertainties in the TF zeropoint and slope are indicated in parentheses. Three separate fits are indicated. In the first (line 1), all objects within any given cluster are assumed to lie at a common distance. In the second fit (line 2), nine clusters have been allowed to "expand" according to the model discussed in the text. The scatter $\sigma$ is computed assuming that the freedom to allow some clusters to expand adds no new degrees of freedom to the model; the scatter $\sigma_s$ assumes that nine extra degrees of freedom have been added. The quantity $N_\sigma$ is the total number of objects involved in the computation of the scatter, which were required to have TF residuals $\leq 1.30$ mag. In practice, all objects the fit met this criterion. The first two fits incorporate the full iterative bias-correction procedure. The third fit (line 3) results from a one-step least squares analysis, with selection bias entirely neglected (but uses the cluster expansion model from line 2).

Parameters of the HMCL Inverse TF Relation

|  | $D\,(\pm)$ | $e\,(\pm)$ | $\sigma_\eta$ | $\sigma_\eta/e$ (mag) | $N_\sigma$ |
|---|---|---|---|---|---|
| No Bias Corr. | $-5.617\,(.029)$ | $0.1201\,(.0025)$ | $0.0481$ | $0.400$ | $346$ |
| Bias Corr. | $-5.579\,(.030)$ | $0.1177\,(.0025)$ | $0.0484$ | $0.411$ | $346$ |

Table 4: The results of fitting an inverse Tully-Fisher relation to the HM sample. The 1-$\sigma$ uncertainties are indicated in parentheses. The 9-cluster expansion model which greatly reduced the scatter in the foward fit has been used here. In the first line of the table, the results are given when the calibration procedure neglects selection bias. In the second line of the table, the results are those obtained following application of the iterative bias correction procedure. The relations given in Table 2 have been used to calculate the $\eta$-dependence of sample selection. The scatter $\sigma_\eta$ has been calculated assuming that nine additional free parameters have been introduced in the model by allowing clusters to expand.

Fit Coefficients for W91

| Quantity | | Coefficient of | | | | | |
|---|---|---|---|---|---|---|---|
| Predicted | const. | $m_r$ | $\eta$ | $A_B$ | $\mathcal{R}$ | $\log(cz) - 3.5$ | $\sigma$ |
| $\log D$(UGC) | 3.724 | 0.197 | 0.316 | 0.000 | 0.413 | 0.000 | 0.119 |
| $m_z$(Zwicky) | -0.145 | 1.087 | 1.016 | 0.827 | 0.458 | 0.000 | 0.413 |

Table 5: Coefficients of the indicated quantities in the linear relations for $\log D_{\mathrm{UGC}}$ and $m_z$ in terms of the TF observables for the W91 sample. Also given are the rms dispersions ($\sigma$) for these relations. The fits were carried out using all W91 galaxies, including W91CL and W91PP. However, the zeropoint of the $m_z$–$m_r$ relation was determined through a fit to the cluster sample only. See text for details.

Parameters of the W91CL TF Relation

|  | $A\,(\pm)$ | $b\,(\pm)$ | $\sigma$ (mag) | $\sigma_s$ (mag) | $N_\sigma$ |
|---|---|---|---|---|---|
| No Expand | $-4.088\,(.048)$ | $7.533\,(.246)$ | $0.412$ | $0.412$ | $155$ |
| Expand 6 | $-4.137\,(.043)$ | $7.703\,(.209)$ | $0.349$ | $0.357$ | $155$ |
| No Bias Corr | $-4.345\,(.042)$ | $7.089\,(.198)$ | $0.331$ | $0.338$ | $155$ |

Table 6: Parameters resulting from fitting a forward Tully-Fisher relation to the W91CL sample. Three separate TF fits are indicated. In the first line ("No Expand"), all clusters are treated as true clusters. In the second and third lines, six clusters are treated as "expanding" according to the model described in the text. The third line gives the parameters resulting from a single-step fit in which no bias corrections have been made.

Parameters of the W91CL Inverse TF Relation

|  | $D\,(\pm)$ | $e\,(\pm)$ | $\sigma_\eta$ | $\sigma_\eta/e$ (mag) | $N_\sigma$ |
|---|---|---|---|---|---|
| No Bias Corr | $-4.218\,(.044)$ | $0.1213\,(.0032)$ | $0.0443$ | $0.366$ | $155$ |
| Bias Corr | $-4.179\,(.044)$ | $0.1193\,(.0032)$ | $0.0445$ | $0.373$ | $155$ |

Table 7: Parameters resulting from fitting an inverse Tully-Fisher relation to the W91CL sample. The 1-$\sigma$ uncertainties are indicated in parentheses. The 6-cluster expansion model which minimized scatter in the foward fit has been used here. In the first line of the table, the results are given when the calibration procedure neglects selection bias. In the second line, the iterative correction procedure of W94 has been applied. In each case, the scatter $\sigma_\eta$ has been computed assuming that there are six additional free parameters in the model, corresponding to the number of clusters allowed to expand. The quantity $\sigma_\eta/e$ is the equivalent magnitude scatter of the inverse TF relation.

Figure Captions

Fig. 1.— The distribution of UGC diameters (top panel) and Zwicky apparent magnitudes (bottom panel), as a function of log redshift, for HM North. The dotted lines show the catalog limits.

Fig. 2.— H I flux characteristics as a function of log redshift, for HM North. The top panel shows the log of the H I flux itself. The bottom panel shows the log of the H I flux-per-channel, given by flux divided by the raw linewidth.

Fig. 3.— Logarithmic UGC diameters (top panel) and Zwicky magnitudes (bottom panel) plotted against the raw I band magnitudes, for HM92 and M93 sample galaxies. In both plots, the photographic quantities plotted have been corrected for the bias due to their observed values being strictly limited. (See text for details.) The dotted lines in each case show the formal catalog limits. In this and later figures, logarithmic diameters are computed as $1 + \log D$ where $D$ is in arcminutes.

Fig. 4.— ESO blue diameters plotted against the logarithm of radial velocity for HM South. The dotted line shows the formal ESO catalog limit. The plot demonstrates that neither HM92 South nor M91 is complete to this limit.

Fig. 5.— H I characteristics for the HM South sample. The upper panel shows the log of H I flux itself, and the bottom panel the log of the H I flux-per-channel.

Fig. 6.— The H I flux-per-channel plotted against the logarithm of the ESO diameter, for galaxies in HM South.

Fig. 7.— Left hand panel: The relation between log ESO diameters and I band apparent magnitudes. The plot depicts HM South galaxies with $\log D_{\mathrm{ESO}} \geq 1.5'$ after they have been fit to $m_I$ and $\mathcal{R}$ only. As in previous figures, the plotted values of $\log D_{\mathrm{ESO}}$ have been corrected for the bias due to the diameter limit, which is shown as a dotted line. The two right hand panels show residuals from this fit with respect to $\eta$ and $\log(cz)$; the trend with $cz$ is significant (see text).

Fig. 8.— Residuals from a linear fit of $\log F_{\mathrm{H\,I}}$ to the corrected I band magnitude. In the upper panel the fit was to the I band magnitude only, and a correlation with log redshift is evident. In the lower panel, a term linear in $\log(cz)$ has been included, resulting in the elimination of the trend of the residuals with redshift.

Fig. 9.— TF residuals vs. log radial velocity for nine clusters in the HM sample. The TF fit has assumed that all galaxies belonging to any one cluster lie at a common distance.

Fig. 10.— TF residuals vs. log radial velocity for the same nine HM clusters as shown in the previous Figure. Now the distances to the individual galaxies in these clusters have been modeled using the cluster expansion model discussed in the text.

Fig. 11.— The forward Tully-Fisher relation for the HM sample. The absolute magnitudes are computed as the bias-corrected apparent magnitudes minus the distance moduli resulting from the fit; the "expanding cluster" model is taken into account when appropriate (see text). The two right hand panels show Tully-Fisher residuals plotted against circular velocity parameter $\eta$.

Fig. 12.— Effect of varying the internal extinction coefficient $C_{int}^{I}$ on the HM TF calibration. The left panel shows how the TF scatter varies with $C_{int}^{I}$. The dotted line corresponds to a 65% confidence interval for $C_{int}^{I}$ (see text). The three panels on the right show the TF residuals as a function of $\mathcal{R}$ for three values of $C_{int}^{I}$. The top panel corresponds to no internal extinction correction, the middle panel to overcorrection, and the bottom panel to the value of $C_{int}^{I}$ ultimately adopted.

Fig. 13.— Logarithmic UGC diameters (top panel) and Zwicky magnitudes (bottom panel) plotted against the raw r band magnitudes for W91 sample galaxies. Both W91CL and W91PP galaxies are shown. In both plots, the photographic quantities plotted have been corrected for the bias due to their observed values being strictly limited. (See text for details.) The dotted lines in each case show the formal catalog limits.

Fig. 14.— Tully-Fisher fit residuals plotted against log radial velocity for six clusters in the W91CL sample. The residuals are those from a fit in which all galaxies in a given cluster are assumed to lie at a common distance.

Fig. 15.— Same as the previous figure, except that the clusters shown have been treated as "expanding" according to the model described in § 3.2.1.

Fig. 16.— Effect of varying the internal extinction coefficient $C_{int}^{r}$ on the W91CL TF calibration. In the left panel the TF scatter is plotted against $C_{int}^{r}$. The dotted line corresponds to a 65% confidence interval for $C_{int}^{r}$ (see text). The three panels on the right show the TF residuals as a function of $\mathcal{R}$ for three values of $C_{int}^{r}$. The top panel corresponds to no internal extinction correction, the middle panel to overcorrection, and the bottom panel to the value of $C_{int}^{r}$ ultimately adopted.

Fig. 17.— The relative biases $\mathcal{B}$ applied to individual galaxies in the HM TF calibration analysis. The actual bias correction is given by $\sigma\mathcal{B}$, where $\sigma$ is the TF scatter. The relative biases are plotted against: (top panel) the TF-predicted raw I band apparent magnitude, $i.e.$, the predicted apparent magnitude $M(\eta) + \mu$ "uncorrected" for extinction and cosmological effects; and (bottom panel) log redshift.

Fig. 18.— Same as the previous figure, except that the relative biases applied to individual galaxies in W91CL are shown.

Fig. 19.— The difference between forward and inverse TF distance moduli for individual clusters in the HM sample, plotted against log redshift. The upper panel shows the results when no bias corrections have been applied; the lower panel shows the results following the application of the full bias-correction procedure.

Fig. 20.— Same as the previous figure, except that the results are shown for W91CL.

Fig. 21.— Differences between the HM and W91CL distance moduli for the ten clusters common to the two samples. The upper panel shows results for the fully bias-corrected forward TF analysis; the lower panel for the uncorrected inverse TF analysis.

Fig. 22.— The differences between individual galaxy distance moduli inferred from the HM TF relation and from the W91CL TF relation for 114 galaxies common to the two samples.